\documentclass[aps, prd, twocolumn, lengthcheck, superscriptaddress]{revtex4-1}
\usepackage{amsfonts, amssymb, amsmath, graphicx, comment, bm, slashed, caption, subcaption, dcolumn,color}
\usepackage{ulem}

\newcommand{\beq}{\begin{eqnarray}}
\newcommand{\eeq}{\end{eqnarray}}
\newcommand{\Lc}{{\cal{L}}}
\newcommand{\bd}{\mathbf}

\newcommand{\calP}{ {\cal P}}

\newcommand{\tr}{\mbox{tr}}

\newcommand{\lqcd}{\Lambda_{ {\rm QCD}} }
\newcommand{\Nc}{N_{ {\rm c}} }
\newcommand{\Nf}{N_{ {\rm f} } }
\newcommand{\rmi}{ {\rm i} }

\newcommand{\la}{\langle}
\newcommand{\ra}{\rangle}

\def\braket#1{\mathinner{\langle{#1}\rangle}}

\begin{document}
\title{Phenomenological QCD equation of state for massive neutron stars}
\author{Toru Kojo}
\affiliation{Department of Physics, University of Illinois at Urbana-Champaign, 1110 W. Green Street, Urbana, Illinois 61801, USA}
\author{Philip D. Powell}
\affiliation{Department of Physics, University of Illinois at Urbana-Champaign, 1110 W. Green Street, Urbana, Illinois 61801, USA}
\affiliation{Lawrence Livermore National Laboratory, 7000 East Ave., Livermore, CA 94550}
\author{Yifan Song}
\affiliation{Department of Physics, University of Illinois at Urbana-Champaign, 1110 W. Green Street, Urbana, Illinois 61801, USA}
\date{\today}
\author{Gordon Baym}
\affiliation{Department of Physics, University of Illinois at Urbana-Champaign, 1110 W. Green Street, Urbana, Illinois 61801, USA}
\affiliation{Theoretical Research Division, Nishina Center, RIKEN, Wako 351-0198, Japan}


\begin{abstract}
We construct an equation of state for massive neutron stars based on quantum chromodynamics phenomenology.   Our primary purpose is to delineate the relevant ingredients of equations of state that simultaneously have the required stiffness and satisfy constraints from thermodynamics and causality.  These ingredients are: (i) a repulsive density-density interaction, universal for all flavors; (ii) the color-magnetic interaction active from low to high densities; (iii) confining effects, which become increasingly important as the baryon density decreases; (iv) nonperturbative gluons, which are not very sensitive to changes of the quark density.   We use the following "3-window" description:
At baryon densities below about twice normal nuclear density, $2n_0$, we use the Akmal-Pandharipande-Ravenhall (APR) equation of state, and at high densities, $\ge (4-7) n_0$, we use the three-flavor Nambu-Jona-Lasinio (NJL) model supplemented by vector and diquark interactions. In the transition density region, we smoothly interpolate the hadronic and quark equations of state in the chemical potential-pressure plane. 
Requiring that the equation of state approach APR at low densities, we find that the quark pressure in {\it nonconfining models} can be larger than the hadronic pressure, unlike in conventional equations of state. We show that consistent equations of state of stiffness sufficient to allow massive neutron stars are reasonably tightly constrained, suggesting  that gluon dynamics remains nonperturbative even at baryon densities $\sim 10\, n_0$. 
 \end{abstract}
\maketitle

\section{Introduction}

\begin{figure*}
\begin{minipage}{1.09\hsize}
\hspace{-1.9cm}
\includegraphics[width = 0.327\textwidth]{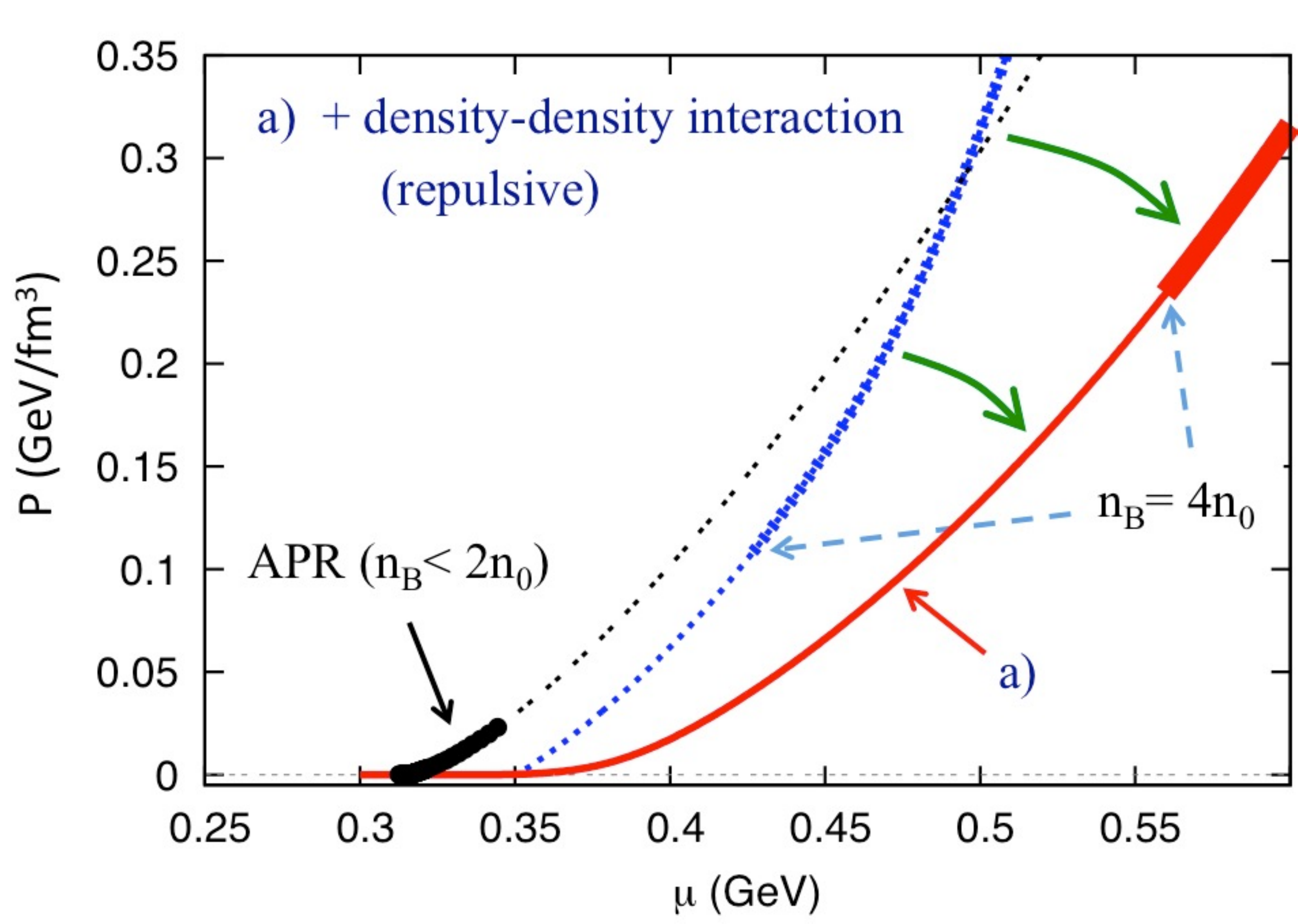}
\hspace{-0.265cm}
\includegraphics[width = 0.3\textwidth]{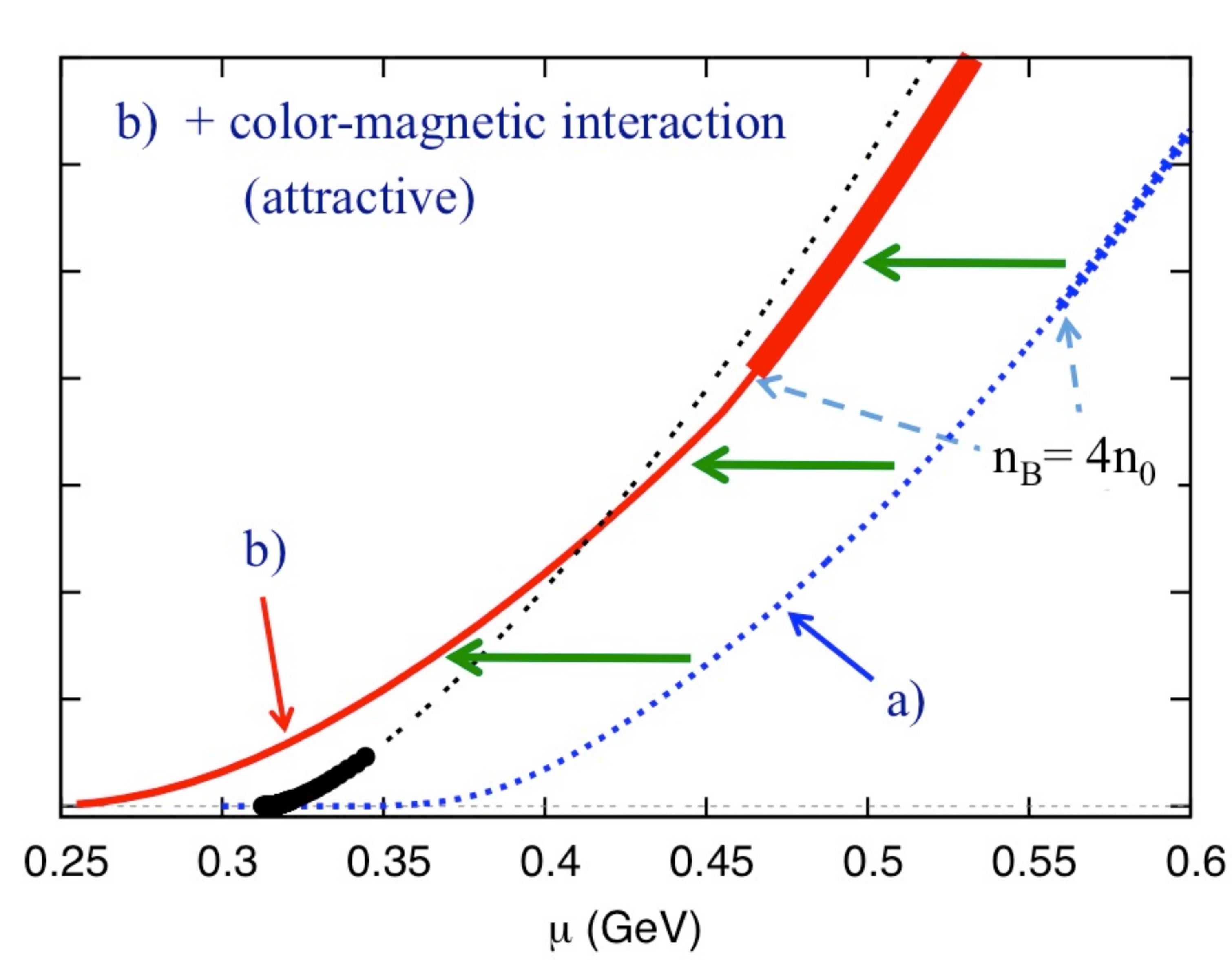}
\hspace{-0.40cm}
\includegraphics[width = 0.3\textwidth]{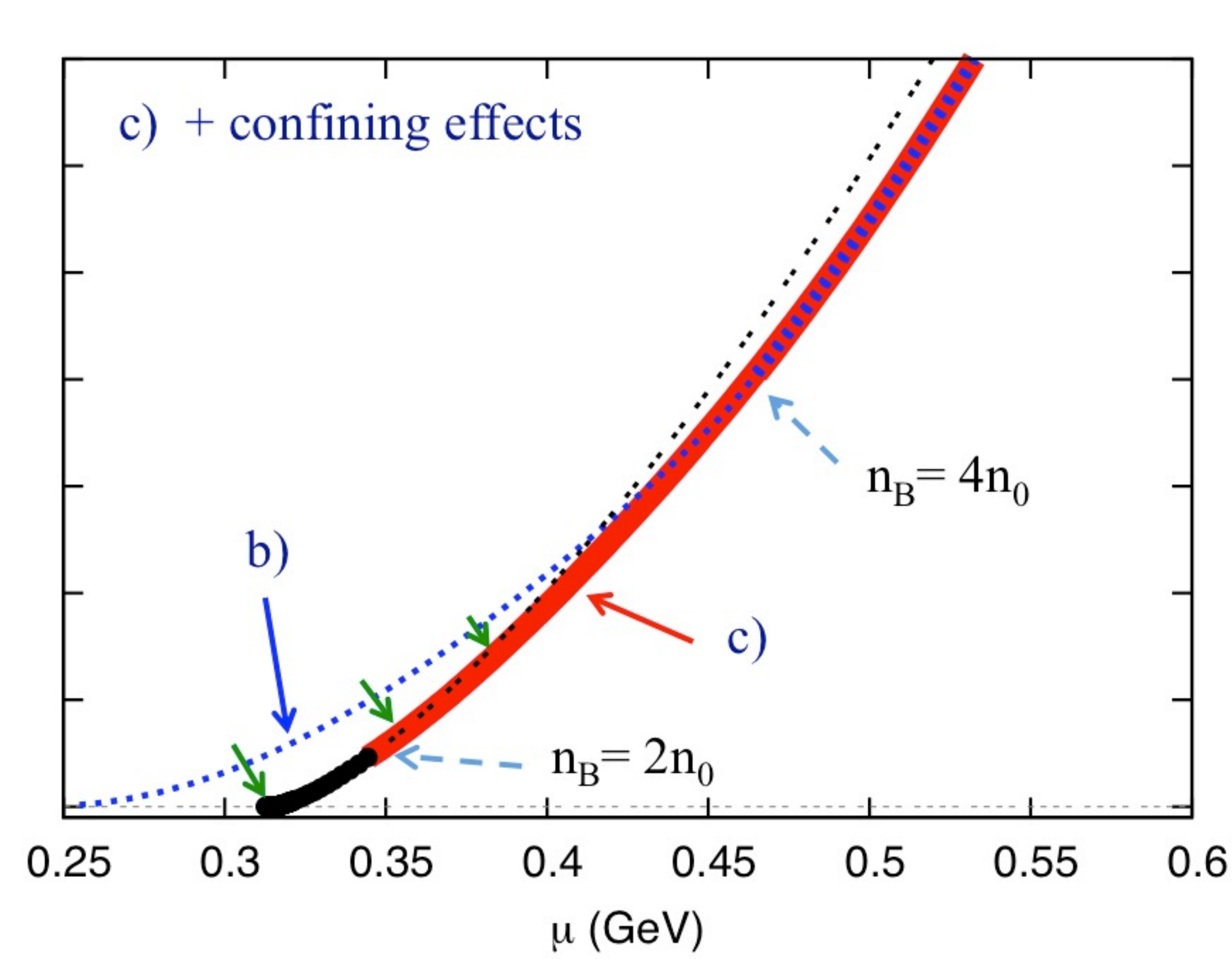}
\end{minipage}
\caption{
\footnotesize{Pressure vs. quark chemical potential for several equations of state.  The black line is the APR result  \cite{Akmal:1998cf} (A18+$\delta v$+UIX* without pion condensation)): the bold line for $n_B<2n_0$ and thin dotted line for $n_B>2n_0$. Various effects are successively added to the standard NJL model; (a) a repulsive density-density interaction, which stiffens the NJL equation of state ; (b) the color-magnetic interaction (diquark correlation), which reduces the average quark energy at all densities; and (c) confining effects, which suppress the artificially large pressure in NJL models at low density down to the APR pressure, discussed in the text. 
}
\vspace{-0.2cm}
}\label{fig:illust}
\end{figure*}

    As the Relativistic Heavy Ion Collider and the Large Hadron Collider continue to push the limits of our experimental knowledge of hot dense quantum chromodynamics (QCD), neutron stars are the only cosmic laboratories in which we can study the structure of cold dense QCD~\cite{BC1976,Lattimer:2006xb,Fukushima:2010bq,Alford:2007xm,Buballa:2014tba}.  The recent discoveries of neutron stars with masses $M \simeq 2 M_\odot$ ($M_\odot$ is the solar mass), including the binary millisecond pulsar J1614-2230 with mass $(1.97\pm0.04)\, M_\odot$ \cite{Demorest} and the pulsar J0348+0432 with mass $(2.01\pm0.04)\, M_\odot$  \cite{Antoniadis2013} (also PSR J1311-3430  \cite{Romani2012}), together with  recent simultaneous determinations of neutron star masses and radii~\cite{Ozel2010,Steiner2012} pose particular challenges to the theoretical construction of the neutron star equation of state.  
    
    On the one hand, the existence of these massive stars suggests that the equation of state must be stiffer than conventional hadronic descriptions of matter including hyperons.  Furthermore, the central baryon density in neutron stars with masses $\sim 2 M_\odot$ well exceeds twice nuclear matter density  $n_0$, and may reach as high as $\sim 10 n_0$.   To understand why such high mass stars are stable requires a knowledge of the equation of state at baryon densities $n_B$ over a range $\sim 1-10 n_0$.  However we cannot at present reliably calculate the equation of state over such a range; the densities are too high to apply reliably conventional hadronic equations of state, and too low to apply perturbative QCD.

    This situation motivates us to investigate the properties of strongly correlated quark matter, intermediate between the hadronic and perturbative QCD phases, and ask how the properties of such matter is constrained by neutron star observations.  Using a schematic quark model, we manifestly take into account quark degrees of freedom, while including interaction effects  such as vector repulsion between quarks, known from hadron spectroscopy, color-magnetic diquark interactions, and six-quark interactions arising from the axial anomaly.   We examine the roles of these interactions and find that it is possible, within a reasonable parameter range, to construct an equation of state that  (i) is sufficiently stiff to include stable stars with $M \sim 2 M_\odot$; (ii) satisfies the thermodynamic constraint that the baryon number density be an increasing function of the baryon chemical potential, $\partial n_B/\partial \mu_B>0$; and (iii) is consistent with the (suggestive) causality constraint that the speed of sound (at zero frequency) not exceed the speed of light \cite{ruderman-bludman,Ellis2007}. While these conditions provide relatively tight constraints on the quark matter equation of state, it is nonetheless possible to construct the desired equation of state using quark model parameters compatible with the hadron spectroscopy.

    To further motivate the picture of strongly correlated quark matter, we briefly review the domain of applicability of hadronic and perturbative QCD equations of state.
Conventional hadronic equations of state are constrained by experimental data at low energy and density, e.g., two-body hadronic scattering below the pion production threshold, the masses and level structure of light nuclei, and nuclear matter around nuclear saturation density $n_0$.  While hadronic equations of state include the relevant physics in the low density regime in which their parameters are fit, with increasing density multiple meson exchanges, many-baryon interactions, and virtual baryonic excitations become increasingly important.  (The systematics can be most clearly seen in the chiral effective theory approach \cite{Weinberg:1978kz,Epelbaum:2008ga,Bogner:2009bt}.)  In nucleonic potential models~\cite{Akmal:1998cf}, the three-body nucleon interaction is crucial to reproducing nuclear matter properties at $n_B\simeq n_0$, and its contribution to the energy density can be even comparable (and of opposite sign) to that of the two-body force at $n_B \sim 2n_0$.  Beyond baryon densities $n_B \gg n_0$ a well defined expansion in terms of static two-, three-, or more, body forces no longer exists.

The equation of state of perturbative QCD~\cite{Freedman:1976ub,Fraga:2001id,Kurkela:2009gj} relies on a picture of weakly coupled quarks and gluons. A current state-of-the-art calculation in this regime, to second order in the strong interaction fine structure constant $\alpha_s$,  with strange quark mass corrections~\cite{Kurkela:2009gj}, finds a relatively strong dependence of the QCD equation of state on the renormalization scale below the quark chemical potential $\mu \sim 1$ GeV, which corresponds to a baryon density $\sim 10^2n_0$.   Such dependence indicates that nonperturbative effects remain quite important in the lower density range relevant to neutron stars.

 In constructing a phenomenological QCD equation of state here, we follow the spirit of the ``3-window" approach of Masuda, Hatsuda, and Takatsuka~\cite{Masuda:2012kf} that interpolates between a nuclear equation of state at low density and a quark equation of state at high density.  At densities below $2 n_0$ we adopt the hadronic Akmal-Pandaripande-Ravenhall (APR) equation of state~\cite{Akmal:1998cf} (denoted in their paper as A18+$\delta v$+UIX*).  At densities above 4-7 $n_0$, where a gas of baryons of radius 0.4-0.5 fm would begin to percolate \cite{perc}, we employ a three-flavor Nambu--Jona-Lasinio (NJL) quark model including vector and diquark interactions. While we use the NJL model to be specific, our discussions are more general. In the intermediate region, where purely hadronic or purely quark descriptions are not appropriate, we construct an equation of state using a smooth polynomial interpolation in the baryon chemical potential--pressure ($\mu$,$P$) plane.  
In this plane the pressure must be a continuous and monotonically increasing function of $\mu$.   One cannot rule out the possibility of a first order transition as the baryon density increases; such a transition would appear as a discontinuity in the first derivative, $\partial P/\partial \mu$.

   The 3-window approach is quite different from the conventional hybrid one in which one regards the quark and hadronic phases as distinct. In the latter, the quark pressure at given $\mu$ must, with increasing $\mu$, intersect the hadronic pressure from below, and moreover must remain larger than the hadronic pressure at larger $\mu$. By regarding the hadronic equation of state at density larger than $\sim 2n_0$ as a valid description of matter, such a hybrid construction implicitly selects out possible forms of quark equations of state;  in order to have an intersection, the quark pressure must be larger than that of the hadronic phase as large $\mu$.  Such quark equations of state are typically soft, and as a consequence hybrid stars with larger quark cores tend to have smaller masses.  In contrast, in the 3-window approach, we construct high density quark equations of state independently of assuming a trustworthy high density hadronic pressure; at high density, the resulting quark pressure at given $\mu$ does not have to grow fast and may remain smaller than the pressure extrapolated from the hadronic phase. Such quark equations of state tend to be stiff, and a star with large quark core can have a large mass.

  Within this schematic 3-window description, we aim to incorporate the following effects known from observed hadronic spectroscopy:  (i) A repulsive flavor-independent density-density interaction \cite{Kunihiro:1991qu}, which stiffens the equation of state (Fig.\ref{fig:illust}a).  (ii) The attractive color-magnetic interaction, relevant at all densities, which reduces the average single quark energy (Fig.\ref{fig:illust}.(b)). This effect is similar to that observed in the constituent quark model~\cite{De Rujula:1975ge}, in which the average quark energy in a nucleon is reduced from the constituent quark mass, $\sim 340$ MeV, to one-third of the nucleon mass, $\sim 313$ MeV.   
As we show, this effect plays an important role in ensuring that the interpolated equation of state satisfies thermodynamic constraints.  (iii) Confinement, which, at low densities, traps quarks into baryons and forbids quarks to contribute significantly to the pressure.  
As we describe, the NJL model, which does not include confinement, has a higher pressure at low density than nuclear models. 
The requirement that the interpolated pressure merges smoothly into APR at low densities (Fig. \ref{fig:illust}c) effectively suppresses such excess pressure. 

The present approach of interpolating between a hadronic and an NJL based quark picture makes the tacit assumption that  the behavior of the gluon sector does not change appreciably over the range $0 \lesssim n_B \lesssim 10 n_0$;   and in particular, that the gluons do not add a bag constant, $B_g$, to the energy (and subtracted from the pressure) when the nonperturbative gluons become perturbative. 
The bag constant measures the energy difference between the trivial (perturbative vacuum) and the nonperturbative vacuum. With the zero-point of the energy set to make the QCD (nonperturbative) vacuum energy zero, the perturbative vacuum has positive energy.  Thus, whenever we consider the extreme conditions under which nonperturbative effects disappear, we must include the bag constant in addition to the contributions of the perturbative effects.
However, the
stability of massive neutron stars does not permit gluon condensation at the QCD scale $\lqcd \sim 0.2$ GeV to produce a gluonic bag constant, $B_g \sim \lqcd^4$, since such a term would, as we argue, too greatly soften the equation of state; we thus exclude the possibility of such a term.  On the other hand, a quark bag constant, $B_q$  of order $\lqcd^4$ associated with restoration of chiral symmetry, is unavoidable in the NJL model~\cite{Asakawa:1989bq}.  

    We extrapolate NJL parameters obtained via hadron phenomenology at $n_B \sim n_0$ to high density quark matter ($n_B \sim 10 n_0$)~\cite{Hatsuda:1994pi,Buballa:2003qv}, an approach that is consistent with the observation from analyses for a large number of colors, $\Nc$, that gluon dynamics is insensitive to quark loop effects~\cite{McLerran:2007qj}. 

This paper is organized as follows. In Sec.~\ref{sec:models}, we briefly describe the hadronic and quark models adopted in this study. In Sec.~\ref{sec:quarkEOS}, we examine interaction effects on the quark equation of state.  In Sec.~\ref{sec:interpolatedEOS}, we  construct the interpolated equation of state. In particular, we explain the difference between the present 3-window description and conventional equations of state which introduce a first order phase transition between hadronic and quark matter in~\cite{BC1976,Alford:2002rj,Alford:2004pf,Klahn:2006iw,Bonanno:2011ch}.  As we see the constraints from thermodynamics and causality are quite important.  In Sec.~\ref{sec:MR}, we solve the Tolman-Oppenheimer-Volkoff (TOV) equation and examine the resulting mass-radius ($M$-$R$) relation of neutron stars. Section~\ref{sec:summary} is devoted to a summary and outlook. 

 We use the following conventions: $g_{\mu \nu}={\rm diag}(1,-1,-1,-1)$, $\gamma_5=\gamma_5^\dag$, and the charge conjugation operator $C$ is $i \gamma_0\gamma_2$. The flavor and color $U(3)$ generator matrices $\tau_i\, (i=0,\cdots, 8)$ and $\lambda_a\, (a=0,\cdots, 8)$ are composed, respectively, of the identity and Gell-Mann matrices, and are normalized as $\tr[\tau_i \tau_j]=2\delta_{ij}$ and $\tr[\lambda_a \lambda_b]=2\delta_{ab}$.  We work in units in which $c$ and $\hbar=1$.

\section{Models}
\label{sec:models}

In this section we briefly summarize the features of the hadronic (APR) and quark (NJL) matter equations of state employed in this paper.  APR will be used to describe low density matter, $n_B<2n_0$, while the NJL model will be used at high densities, $n_B > (4-7)n_0$.  The precise density beyond which we adopt a fully quark description of matter will depend upon details of the interpolation, as discussed in section~\ref{sec:interpolatedEOS}.

\subsection{The APR equation of state}
\label{sec: APR}

In this work we adopt the A18+$\delta v$+UIX$^*$ version of the APR equation of state to describe low density hadronic matter~\cite{Akmal:1998cf}.  This equation of state, based on the Argonne $v_{18}$ two-body potential, which fits hadronic scattering data very well, and the Urbana IX three-body interaction, which is important to explain nuclear saturation properties,  includes charge neutrality and $\beta$-equilibrium.  The $\delta v$ indicates the inclusion of relativistic corrections.  For simplicity, we adopt the APR version excluding neutral pion condensate, which emerges at $n_B \sim 1.4 n_0$.  
The APR model includes only nucleonic degrees of freedom, and does not take into account hyperons, whose interactions with nucleons and among themselves are not well determined. Typical models of nucleon-hyperon interactions predict hyperon onset at a density $n_B \sim 2-3 n_0$.   We restrict our application of  APR to $n_B< 2n_0$.

\subsection{The NJL equation of state}
\label{sec: NJLmodel}

\subsubsection{The Lagrangian}

In descriptions of quark matter, we adopt a three flavor Nambu--Jona-Lasinio model with Lagrangian density
\beq
\Lc = \overline{q} (\rmi \slashed{\partial} - \hat{m}) q + \Lc^{(4)} + \Lc^{(6)}   ,   \label{eq:NJL_Lagrangian}
\eeq
where $q$ is a quark field with color, flavor, and Dirac indices, $\hat{m}$ is the quark current mass matrix, and $\Lc^{(4)} = \Lc^{(4)}_\sigma + \Lc^{(4)}_V + \Lc^{(4)}_d$ and $\Lc^{(6)} = \Lc^{(6)}_\sigma + \Lc^{(6)}_{\sigma d}$ are four- and six-quark interaction terms, respectively, chosen to reflect the symmetries of QCD. The four-quark interactions possess $U_L(3) \times U_R(3)$ symmetry for flavors, while the six-quark interactions reflect the $U_A(1)$ anomaly.

The first of the four-quark interactions describes spontaneous chiral symmetry breaking:
\beq
\Lc^{(4)}_\sigma = G \sum^8_{i=0} \left[ (\overline{q} \tau_i q)^2 + (\overline{q} \rmi \gamma_5 \tau_i q)^2 \right] = 8 G \mbox{tr} (\phi^\dagger \phi)   ,   
\label{eq:L4_sigma}
\eeq
where $G > 0$ (attractive), and $\phi_{ij} = (\overline{q}_R)^j_a (q_L)^i_a$ is the chiral operator with flavor indices $i,j$ (summed over the color index $a$).  

The second of the four-quark terms~\cite{Kitazawa:2002bc},
\beq
\Lc^{(4)}_V = - g_V (\overline{q} \gamma^\mu q)^2   ,   ~~~~~(g_V>0)
\label{eq:L4_V}
\eeq
describes the repulsive density-density interaction, analogous to $\omega$-meson exchange in nuclear matter.

The third of the four-quark terms,
\beq
\Lc^{(4)}_d & = & H \sum_{A,A^\prime = 2,5,7} \big[ \left(\overline{q} \rmi \gamma_5 \tau_A \lambda_{A^\prime} C \overline{q}^T \right) \left(q^T C \rmi \gamma_5 \tau_A \lambda_{A^\prime} q \right)   \nonumber   \\
	       && \hspace{23mm} + \left(\overline{q} \tau_A \lambda_{A^\prime} C \overline{q}^T \right) \left(q^T C \tau_A \lambda_{A^\prime} q \right) \big]   ,   \nonumber   \\
	    & = & 2 H \tr (d^\dagger_L d_L + d^\dagger_R d_R)   \label{eq:L4_d}\,,   ~~~~~(H>0),
\eeq
describes attractive diquark pairing, where $\tau_A$ and $\lambda_{A^\prime}$ ($A,A'=2,5,7$) are the antisymmetric generators of U(3) flavor and SU(3) color, respectively.  The structure of the interaction can be understood as the color-magnetic interaction in the $2\rightarrow2$ scatterings of quarks in $s$-wave, spin-singlet, flavor- and color- anti-triplet channel. The operators $(d_{L,R})_{ai} = \epsilon_{abc} \epsilon_{ijk} (q_{L,R})^j_b C (q_{L,R})^k_c$ are diquark operators of left- and right-handed chirality.

Next we discuss the six-quark interactions responsible for the $U_A(1)$ anomaly~\cite{Kobayashi:1970ji}.  The first term involves the product of the chiral condensates of different flavors:
\beq
\Lc^{(6)}_\sigma & = & -8 K ( \det\,\!\!_{\rm f} \phi + \mbox{h.c.}) \,,   ~~~~(K>0)
\eeq
where $\det_{\rm f}$ denotes the determinant with respect to flavor indices.  The second term couples the chiral and diquark condensates~\cite{Hatsuda:2006ps},
\beq
\Lc^{(6)}_{\sigma d} & = & K^\prime ( \tr [(d^\dagger_R d_L) \phi] + \mbox{h.c.})\,,~~~~(K'>0)   .
\eeq
At tree level, these two interactions may be related via a Fierz transformation, which leads to the conclusion $K^\prime = K$.  However, renormalization effects will, in general, destroy this equality so that at the mean field level we may treat $K$ and $K^\prime$ as independent parameters, but with $K^\prime \sim K$.

\subsubsection{Electric and color charge neutrality constraints}

In order to avoid energetically expensive static long-range electric Coulomb interactions and color flux tube configurations in stable homogeneous  quark matter, we impose the local electric and color charge neutrality constraints
\beq
n_Q (x)= n_{a} (x) = 0\,. ~~~~(a=1,\cdots,8)
\eeq
where $n_Q$ is the local electric charge density, and the $n_a$ are the local color densities.  These conditions are enforced via standard Lagrange multipliers -- with the appropriate chemical potentials coupled to the electric and color charge densities, respectively~\cite{Iida:2000ha}.

  Introducing the charge chemical potential $\mu_Q$, we add to the Lagrangian the terms
\beq
\Lc_Q = \mu_Q (q^\dagger Q q - l_i^\dagger l_i ) \,,~~~(i: {\rm summed})
\eeq
where $Q = \mbox{diag} (2/3,-1/3,-1/3)$ is the quark charge operator for $(u,d,s)$ quarks, in units of the proton charge $e$. The $l_i =(e,\mu,\tau)$ are lepton fields and $m_i$ the lepton masses.  We may safely omit contributions from $\mu$- and $\tau$- leptons since their populations are vanishingly small in the density range of interest \cite{Masuda:2012kf}.

Colors and flavors in dense matter are coupled through the diquark interactions of $\Lc^{(4)}_d$.  Thus, an asymmetry in quark flavor densities (e.g., a 2SC phase) leads to a corresponding net quark color density.  Most generally case, we should introduce eight independent color chemical potentials~\cite{Buballa:2005bv}.  However, for the diquark pairing structures considered in this paper, all color densities except $n_3=\la q^\dag \lambda_3 q \ra $ and $n_8=\la q^\dag \lambda_8 q \ra$ automatically vanish~\cite{Abuki:2005ms}.  Thus, we only need to add the terms
\beq
\Lc_{3,8} = \mu_3 q^\dagger \lambda_3 q + \mu_8 q^\dagger \lambda_8 q\, ,
\eeq
to constrain the system. The values of $(\mu_Q, \mu_3,\mu_8)$ will be tuned to satisfy the neutrality conditions.

\subsubsection{Mean field equation of state}

The mean fields for the chiral condensate and quark densities are
\beq
\sigma_i & = & \braket{\overline{q}_i q_i}   ,  \hspace{3mm}  \hspace{3mm}  n =\sum_{i=1}^3 \braket{q^\dagger_i q_i} \,.
\eeq
Below we write $(\sigma_1, \sigma_2, \sigma_3)=(\sigma_u, \sigma_d, \sigma_s)$ for later convenience.  For the diquark mean fields, we write
\beq
d_i &= \braket{q^T C \rmi \gamma_5 R_i q} \,,
\eeq
where 
\beq
\left( R_1, R_2, R_3 \right) \equiv \left( \tau_7 \lambda_7, \tau_5 \lambda_5, \tau_2 \lambda_2 \right) \,.
\eeq
With these definitions, the diquark condensates $(d_1, d_2, d_3)$ correspond to $(ds, su, ud)$ quark pairings, respectively.

The thermodynamic potential may be computed from the mean field particle propagators in terms of these mean fields; the inverse of the propagator $S(k)$, can be read off from the mean field Lagrangian \cite{Hatsuda:2006ps}, 
\beq
S^{-1} (k) = \begin{pmatrix}
		~ \slashed{k} - \hat{M} + \hat{\mu} \gamma^0 ~&~ \rmi \gamma_5 \Delta_i R_i ~\\
		~ - \rmi \gamma_5 \Delta^\ast_i R_i ~&~ \slashed{k} - \hat{M} - \hat{\mu} \gamma^0 ~
	     \end{pmatrix}   ,   \label{eq:Sinv}
\eeq
where the effective mass matrix has diagonal elements
\beq
M_i = m_i - 4 G \sigma_i + K |\epsilon_{ijk}| \sigma_j \sigma_k + \frac{K^\prime}{4} \hspace{.5mm} |d_{i}|^2   ,
\eeq
while the three diquark pairing amplitudes,
\beq
\Delta_i = -2 d_i \left(H - \frac{K^\prime}{4} \hspace{.5mm} \sigma_i \right)   ,
\eeq
and the effective chemical potential matrix,
\beq
\hat{\mu} = \mu - 2 g_V n + \mu_8 \lambda_8 + \mu_Q Q \,,   \label{eq:mu_eff}
\eeq
are color- and flavor-dependent.

For each momentum, the inverse propagator is a $72\times 72$ matrix. There is spin degeneracy, so maximally there are $36$-independent eigenstates in the presence of the flavor and color asymmetry.  In the Nambu-Gor'kov formalism, the eigenenergies appear as pairs, $(\epsilon,-\epsilon)$. The single particle contribution to the thermodynamic potential is 
%
%
%
\beq
\Omega_{ {\rm single} } & = & - 2 \sum^{18}_{j=1} \int^\Lambda \!\!\frac{d^3 \bd{k}}{(2\pi)^3} 
\left[ T \ln \left( 1 + e^{- |\epsilon_j | /T }  \right)+ \frac{ \Delta \epsilon_j}{2} \right] \,,\nonumber \\
\eeq
where $\Delta \epsilon_j = \epsilon_j -\epsilon_j ^{\rm free}$; here $\Lambda$ is an ultraviolet cutoff. The $\mu$-dependence is hidden in the eigenvalue $\epsilon_j$. Because Eq. (\ref{eq:Sinv}) cannot be inverted analytically, the eigenvalues must be computed numerically for each momentum~\cite{Blaschke:2005uj}.

In order to remove the double-counting of interactions typical of mean-field treatments, we must also include in the thermodynamic potential the terms
\beq
\Omega_{ {\rm cond} } & = & \sum^3_{i=1} 
\left[ 2G\sigma^2_i + \left(H - \frac{K^\prime}{2} \, \sigma_i \right) |d_i|^2 \right]\nonumber \\
&& \hspace{10mm} - 4 K \sigma_1 \sigma_2 \sigma_3 - g_V n^2 \,.
\eeq
These terms are positive ($\sigma_i <0$), except for the final term.

The quark matter thermodynamic potential is $\Omega^{ {\rm bare} }_{q} = \Omega_{ {\rm single} } + \Omega_{ {\rm cond.} }$.  However, there still remains the nontrivial choice of the ``zero" of the thermodyanmic potential.  For discussions of neutron star masses, this procedure is extremely important because in general relativitity, the absolute energy density, as in the TOV equation,  and not simply its deviation from the QCD vacuum, is physically relevant.   Given that the cosmological constant is extremely small compared to the QCD scale, we set the origin of the thermodynamic potential to zero at zero quark density and temperature.  Thus in constructing the quark matter equation of state, we will use the renormalized thermodynamic potential
\beq
\Omega_q (\hat{\mu},T) \equiv \Omega^{{\rm bare}}_{q } (\hat{\mu},T) -\Omega^{{\rm bare}}_{q } (\hat{\mu}=T=0) \,,
\eeq
which vanishes at $T=\hat{\mu}=0$.

Finally, the electron contribution to the thermodynamic potential is the standard
\beq
\hspace{-0.3cm}
\Omega_e = -2 T\sum_{\lambda=\pm}  \int \frac{d^3 \bd{k}}{(2\pi)^3} \,\ln \left(1 + e^{-(E_{e} + \lambda \mu_Q)/T} \right) \,,
\eeq
with $E_{e} = \sqrt{\bd{k}^2 + m^2_e}$, and where we recall that the electron chemical potential is $\mu_{e^-} = - \mu_Q$.

Writing the total thermodynamic potential as $\Omega = \Omega_q + \Omega_e$, the thermodynamic state of the system is determined minimizing the free energy with respect to the seven condensates \{$\sigma_i$,$d_i$,$n$\} under the neutrality conditions
\beq
n_{Q, 3, 8} = - \frac{\, \partial \Omega \,}{\, \partial \mu_{Q, 3, 8} \,}  = 0 ,  \label{eq:constraint_1}
\eeq
which yields the ``gap equations."
\beq
0 = - \frac{\, \partial \Omega \,}{\, \partial \sigma_i \,} = -\frac{\, \partial \Omega \,}{\, \partial d_i \,} \,,
~~~~ n = -  \frac{\, \partial \Omega \,}{\, \partial \mu \,} \,.
\eeq
Below we solve these self-consistent equations using the method outlined in~\cite{Abuki:2005ms}. Whenever we encounter regions in the solution suggestive of first order phase transitions, we explicitly compare $\Omega$ in the relevant phases to determine which local minima is gives the lower free energy.  For the ground state we calculate at a nonzero but very small temperature $\sim 0.1 m_e$, which makes the numerical calculations faster and more stable. 

\subsubsection{The NJL parameters}

For the model outlined above, we identify two distinct sets of parameters: $(\Lambda, m_{u,d}, m_s, G, K)$ and $(g_V, H, K')$. The first set is fixed by matching to QCD vacuum phenomenology.  In this work we will use the set by Hatsuda and Kunihiro (HK)\cite{Hatsuda:1994pi} (Table~\ref{tab:couplings}), which gives the vacuum effective masses for light flavors, $M_{u,d} \simeq 336$ MeV, and strange quark, $M_s\simeq 528$ MeV.  The second set of parameters does not manifestly affect the quantities in QCD vacuum at the mean field level; we therefore treat them as free parameters, but of the same order of magnitude as the first set, based upon Fierz transformations connecting the associated interaction vertices in the absence of any known anomalous suppression.
Briefly, we will investigate $g_V=1-2G$, $H=1-1.5G$, and $K'=0$ and $K'=K$. The choice of these values will be explained in Sec.\ref{qualitative}.

\begin{center}
\begin{table}[t]
\caption{\footnotesize{Three common parameter sets for the three-flavor NJL model: the average up and down bare quark mass $m_{u,d}$, strange bare quark mass $m_s$, coupling constants $G$ and $K^\prime$, and 3-momentum cutoff $\Lambda$~\cite{Hatsuda:1994pi,Rehberg,Lutz1992}, with $\Lambda$, $m_{u,d}$, and $m_s$ in MeV.}}
\begin{tabular}{|c|c|c|c|c|c|}
\hline 
   & $\Lambda$ & \ $m_{u,d}$ & \ $m_s$ & \ $G \Lambda^2$ \ & \ $K \Lambda^5$ \ \\ \hline \hline
HK  \cite{Hatsuda:1994pi}& 631.4 & 5.5 & 135.7 & 1.835 & 9.29 \\ \hline
RHK \cite{Rehberg} & 602.3 & 5.5 & 140.7 & 1.835 & 12.36 \\ \hline
LKW \cite{Lutz1992} & 750.0 & 3.6 & 87.0 & 1.820 & 8.90 \\ \hline
\end{tabular}
\label{tab:couplings}
\end{table}
\end{center}

\section{Qualitative effects on the quark equation of state}
\label{qualitative}

In this section we examine a number of qualitative effects related to the quark matter equation of state.

\subsection{When do the quark equations of state become stiff? \label{stiffEOS}}

\label{sec:quarkEOS}

We begin our discussion of stiffening of the equation of state by considering a schematic expression for the quark sector equation of state (see also~\cite{Alford:2004pf}). For simplicity, we presently consider only matter in a single phase, ignoring any complications arising from phase transitions.  In this context, the energy density may be parameterized in terms of the quark density as
\beq
\varepsilon(n) = c_1 n^{4/3} + c_2 n^{2/3} + c_{-2} n^2 + B \,,      \label{eq:energy}
\eeq
where $n \sim p_F^3$ with $p_F$ the quark Fermi momentum.  The first term is the kinetic energy contribution.  The second term contains contributions from both diquark pairing on the Fermi surface ($\sim - p^2_F \Delta(p_F)^2$) and mass corrections ($\sim + p^2_F M(p_F)^2$), where we assume that $\Delta,M \ll p_F$.  (The limit $M \ll p_F$ is applicable for all quarks, even strange, 
at high density, where chiral restoration occurs, and is applicable for u and d quarks at intermediate density.)  The third term represents the density-density interaction. The last term is the bag constant $B\,(>0)$.  We neglect, for large $p_F$, the density dependence of the pairing gaps, as well as that of the bag constant.

Differentiating (\ref{eq:energy}) yields the chemical potential $\mu=\partial \varepsilon/\partial n$, from which the pressure $P=\mu n-\varepsilon$ is:
\beq
P = \frac{\, \varepsilon \,}{3} - \frac{2}{\, 3\,}c_2 n^{2/3} + \frac{2}{\, 3 \,}c_{-2} n^{2} - \frac{\,4\,}{3}    B \,.
\eeq
The first term is the kinetic pressure, while the remaining terms correspond to corrections arising from the mechanisms discussed above.  For given $\varepsilon$, the pressure becomes large when $c_2 < 0$ and $c_{-2} > 0$.  The former condition is met when the quarks interact attractively near the Fermi surface.  The latter condition simply expresses the requirement that the density-density interaction should be repulsive in order to stiffen the equation of state.  More generally, for stiff equations of state, the coefficients $c_{m\ge 2}$ should be negative, while $c_{m<0}$ should be positive.  Finally, a smaller value of the bag constant also tends to stiffen the equation of state.

\subsection{Quark and gluon bag constants: $B_q$ and $B_g$    \label{bagconstants}}

To consider the impact of the quark and gluon bag constants, we begin by supposing that both the quark and gluon sectors are weakly interacting and that all quark and gluon condensates are vanishingly small.  In the absence of perturbative corrections, the equation of state is then %
\beq
P(\mu) = c_0 \mu^4 - B \,,~~~~~\varepsilon(\mu) = 3 c_0 \mu^4 + B \,,
\eeq
where $c_0 = \Nc \Nf/12\pi^2$ is a function of the number of quark colors $\Nc$ and flavors $\Nf$, and the net bag constant is sum of quark and gluon contributions, $B = B_q + B_g$. 

The existence of the bag constant changes the energy-pressure relation from  $\varepsilon = 3 P$ to $\varepsilon - 4B = 3P$. Therefore, a smaller $B$ enhances $P$ at given $\varepsilon$, stiffening the equation of state.  In fact, for a three flavor ideal quark gas with a bag constant, the maximum neutron star mass scales as~\cite{Book,Witten:1984rs} 
\beq
M_{ {\rm max} } \simeq 1.78\,M_\odot \left(\frac{155\, {\rm MeV} }{B^{1/4} }\right)^2   ,
\label{idealgasM}
\eeq
while the corresponding radius scales as
\beq
R \simeq 9.5\, \left(\frac{155\, {\rm MeV} }{B^{1/4} }\right)^2 \, {\rm km}   .
\eeq
Thus, smaller values of $B$ give rise to more massive, larger neutron stars.

In the NJL model, the quark bag constant at large density appears automatically when the gap in the Dirac sea is closed through chiral restoration. As a result, its value be computed explicitly as
\beq
B_q^{ {\rm NJL} } \equiv
\left[\, \Omega(M_{ {\rm eff} } = m) - \Omega(M_{ {\rm eff} } =M) \,\right]_{T=\mu=0} ,
\eeq
where $M_{ {\rm eff} }$ is the effective mass in the quark energy; $M_{ {\rm eff} }$ becomes the current quark mass ($m$) in the perturbative vacuum and the dynamically generated mass ($M$) in the chiral symmetry-broken vacuum.
In the HK parameter set with vacuum effective masses $M_{u,d}=336\, {\rm MeV}$ and $M_{s}=528\, {\rm MeV}$, the bag constant is
\beq
B_q^{ {\rm NJL} } \simeq 284\, {\rm MeV/fm^3} = (219\, {\rm MeV})^4 \,.
\label{bnjl}
\eeq 
Naively substituting this value into Eq.~(\ref{idealgasM}), we obtain a maximum neutron star mass $\sim 0.9 M_\odot$, which less than 1/2 the mass of observed massive stars.  This low maximum mass indicates the importance of interaction effects in order to sustain massive neutron stars.   
  
  Typical values of the bag constant used previously \cite{Book,Akmal:1998cf} are in the range $B^{1/4} \sim 150 -200$ MeV.   The bag constant used in \cite{Book} to construct strange quark stars, $B^{{\rm sq}} \simeq (155{\rm MeV})^4$, is one-fourth of the NJL bag constant, $B^{{\rm sq}} \simeq B_q^{ {\rm NJL} }/4$. For a three-flavor free quark gas with $B^{{\rm sq}}$, the star mass is relatively large, $\simeq 1.78M_\odot$, but still does not reach $\sim 2M_\odot$; to do so requires including interactions.
 Since the value of the bag constant is not precisely known, we employ the NJL value (\ref{bnjl}) for consistency. 
As noted, we should also consider a gluon bag constant, $B_g\sim \lqcd^4$ when gluons become perturbative at some large quark density. However, because the quark bag constant in the NJL model alone already provides considerable softening of the equation of state, a significant contribution from the gluonic bag constant is unlikely in our equation of state, as we show later in Sec.~IV C.  

  One might argue that considering a gluonic bag constant is unnecessary because the gluons are integrated out in determining the interactions in the NJL model, and thus the quark bag constant already contains the gluonic contributions. This is not quite correct.  In the NJL model, the long-range components of the gluons such as those producing confinement are certainly not taken into account; integrating out the long-range components generally produces nonlocal interactions among quarks, which are  not present in the NJL model. Therefore, we must consider the contributions from long-range gluons separately, and not simply ignore the gluonic bag constant. 

\subsection{Repulsive density-density interaction: $g_V$}

\begin{figure}
\includegraphics[width = 0.45\textwidth]{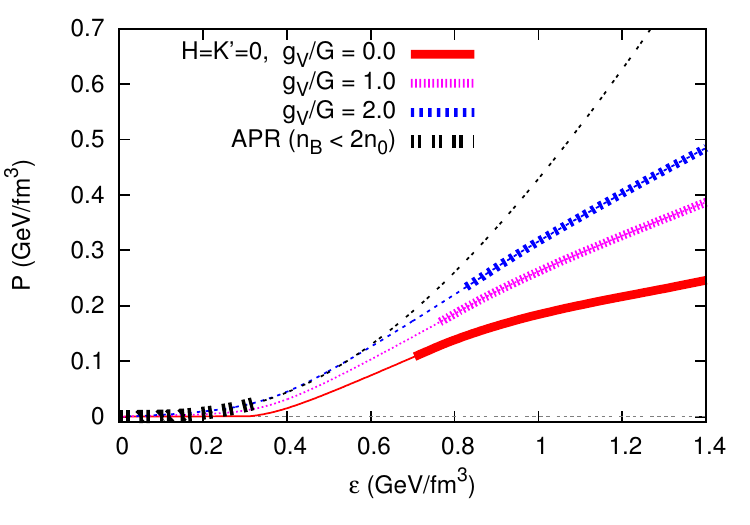}
\caption{\footnotesize{Pressure vs. energy density for several NJL parameter sets and APR.  The bold lines for the NJL equations of state indicate that $n_B > 4 n_0$.  As $g_V$ increases, the NJL equation of state becomes noticeably stiffer.   APR is also plotted for comparison, in bold for $n_B < 2n_0$, and as double dots in the region above, where we do not use APR.}}
\label{fig:gV_ep}
\vspace{-0.2cm}
\end{figure}

\begin{figure}
\begin{minipage}{1.0\hsize}
\includegraphics[width = 0.8\textwidth]{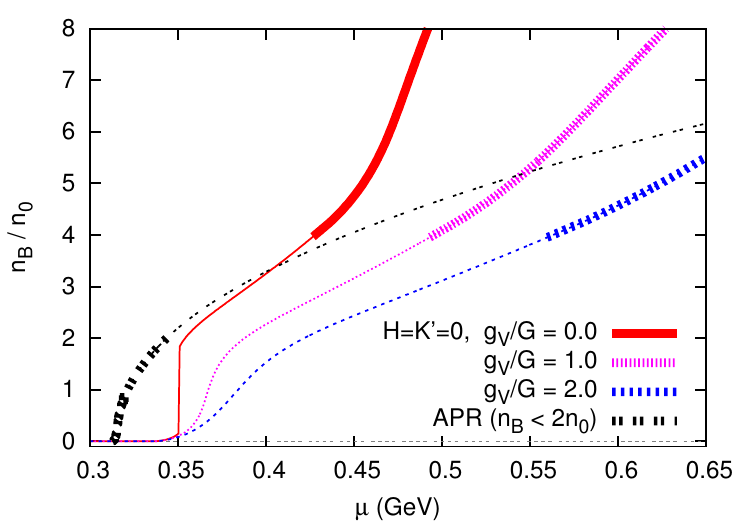}
\end{minipage}\\
\begin{minipage}{1.0\hsize}
\includegraphics[width = 0.85\textwidth]{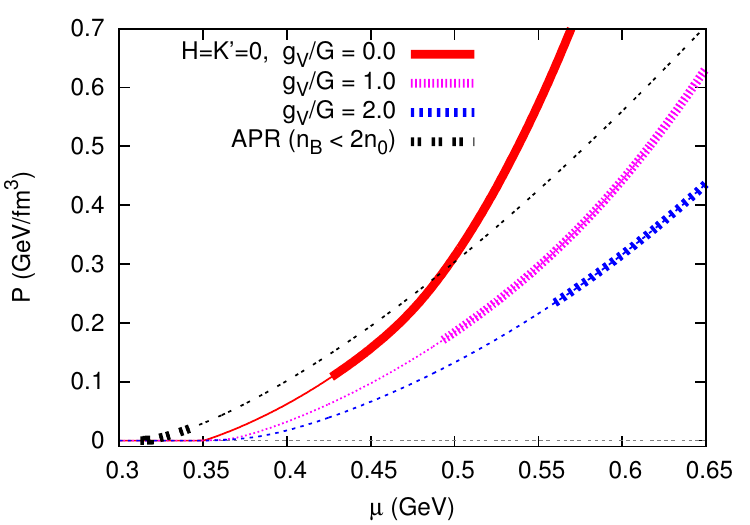}
\end{minipage}
\caption{
\footnotesize{Normalized baryon density $n_B/n_0$ (top panel) and pressure $P$ (bottom panel) as a function of quark chemical potential $\mu$ for several NJL parameter sets and APR.  Typically, the baryon density and pressure in the APR rise faster at lower chemical potential than in  NJL which has a larger effective quark mass, $M \simeq 336$ MeV, than 1/3 of the nucleon mass.}
}\label{fig:mu_np_H0}
\vspace{-0.2cm}
\end{figure}

   The repulsive quark vector interaction is inspired by the repulsive density-density interaction in nuclear matter, described, e.g., by omega meson exchange \cite{Walecka:1974qa}.  Extrapolating the picture of nuclear matter to the strongly correlated quark matter domain, we anticipate that the quark vector coupling is of similar magnitude to the hadronic coupling scale $g_V \sim G$.   
   
    Reference~\cite{Bratovic:2012qs} demonstrated that the vector coupling should be $g_V \sim 2G$ in order to explain the lattice results on the curvature of the chiral restoration line near zero density~\cite{Kaczmarek:2011zz}. (Note that the coupling constant $G$ here corresponds to half that used in Ref. \cite{Bratovic:2012qs}.)  In the following, we focus mainly on the value $g_V = 2G$.

  The inclusion of a vector coupling smooths out chiral symmetry restoration~\cite{Bratovic:2012qs} because the density-density repulsion forbids a rapid increase of the baryon density, and as a result the chiral transition also does not occur rapidly.  Indeed, beyond a particular critical coupling, a first order chiral transition is turned into a smooth crossover. 

Intuitively, an increasing repulsive vector force stiffens the equation of state, $P$ vs. $\varepsilon$, as shown in Fig.~\ref{fig:gV_ep}.  While the NJL equation of state is considerably softer than APR for small vector couplings, when $g_V$ is sufficiently large the NJL equation of state can achieve a stiffness on par with APR across a wide range of densities.    By increasing $g_V$ sufficiently, we can obtain an equation of state within the present framework stiff enough to support neutron star masses $\sim 2M_\odot$.

 On the other hand, increasing $g_V$ makes it more difficult to interpolate between the APR and NJL regimes. This challenge is seen in the plots of $n_B (\mu)$ and $P (\mu)$ in Fig.~\ref{fig:mu_np_H0},  where for both $P$ and $n_B$, the APR and NJL curves become more widely separated in $\mu$ as $g_V$ increases.   One might imagine that the matching could be performed rather simply by allowing  a first order phase transition in the interpolated region; however a first order transition, a sudden increase in $n_B$ is simply a kink in $P$ vs. $\mu$, which does not help the interpolation.
 
  A part of the difficulty of interpolating between APR and NJL is that the
constituent quark mass for light flavors is $M_{u,d} \simeq 336$ MeV,  larger than the one-third of the nucleon mass.    Accordingly, the $P (\mu)$ curve in the NJL model tends be below that of  APR.    We next discuss two-body correlations mediated by the color magnetic interaction, which tend to shift the $P (\mu)$ curves to the left, rendering the interpolation procedure more feasible.


\subsection{Two-body correlations: the color magnetic interaction $H$}
\label{sub:H}

\begin{figure}
\includegraphics[width = 0.45\textwidth]{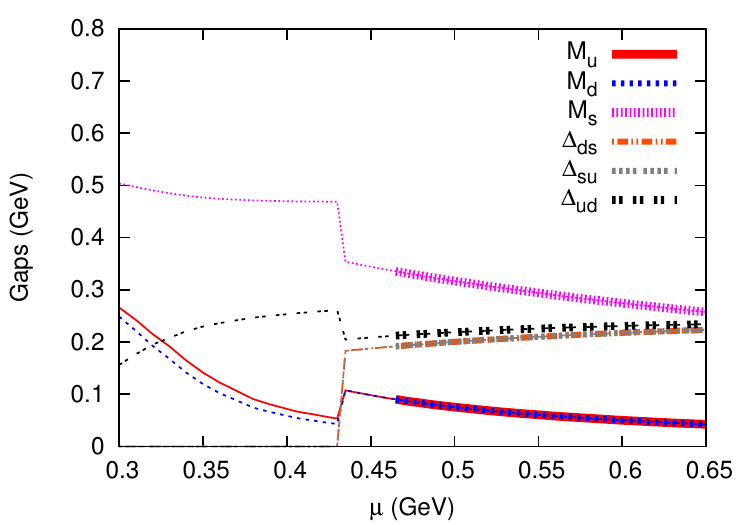}
\caption{\footnotesize{Chiral and diquark gaps as a function of quark chemical potential for $K'=0$, $g_V = 2.0 G$, $H = 1.5 G$.   With increasing density, the system undergoes a first order phase transition from the 2SC to the CFL phase.  The gaps are shown as bold lines for $n_B>4n_0$. }}
\label{fig:mu_con_H}
\vspace{-0.2cm}
\end{figure}

\begin{figure}
\begin{minipage}{1.0\hsize}
\includegraphics[width = 0.9\textwidth]{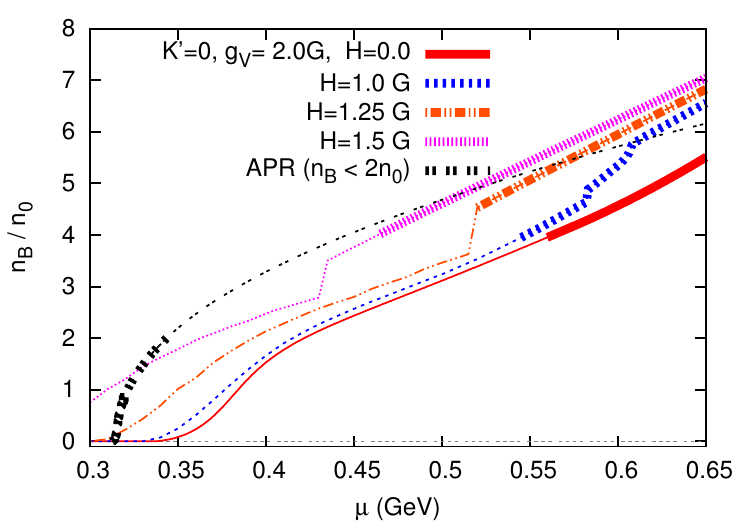}
\end{minipage}\\
\begin{minipage}{1.0\hsize}
\includegraphics[width = 0.95\textwidth]{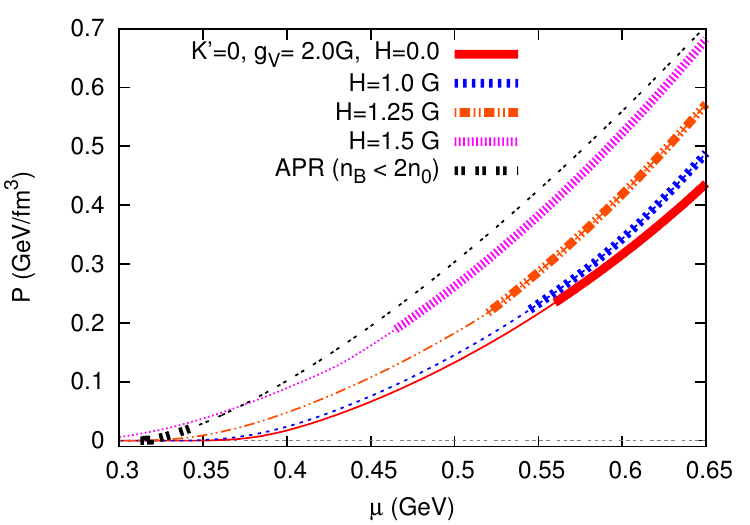}
\end{minipage}
\caption{
\footnotesize{Normalized baryon density $n_B/n_0$ (top panel) and pressure $P$ (bottom panel) as functions of quark chemical potential $\mu$.   We take the vector and axial anomaly couplings in NJL to be $g_V/G=2.0$ and $K^\prime = 0$, and allow the diquark coupling to take on the values $H/G$ = 0, 1.0, 1.25, and 1.5.  As $H$ increases the curves are shifted toward lower chemical potential.}
}\label{fig:mu_np_H}
\vspace{-0.2cm}
\end{figure}

At high density, quarks undergo BCS pairing (diquark condensation) as a consequence of the color magnetic interaction.  Pairing reduces
the energy density by an amount $\delta \varepsilon \sim - p^2_F \Delta^2$, or equivalently, enhances the pressure by $\delta P \sim + p^2_F \Delta^2$. 

   We expect correlation effects among the quarks to increase with decreasing density. Eventually three-quark correlations must be dominant in the hadronic phase. One path to three-quark correlations is increasing diquark correlations plus diquark-single quark correlations beyond that described by the standard choice of the diquark coupling $H$ = 0.75-1.0 $G$, based on Fierz transformation of the one-gluon exchange type vertex.  In this range, we do not find significant effects of $H$ on the equation of state in the density range of interest.  A diquark mean field appears only when the Fermi surface becomes sufficiently large to overcome chiral symmetry breaking effects.   However, diquark correlations, which can exist even without a large Fermi sea, reduce the energy of a pair to less than twice the effective quark mass.   
 To simulate such effects within the present mean field approach, we allow the diquark mean field to reduce the single quark energy at all densities by exploring somewhat larger values $H$ = 1.0-1.5 $G$ than the standard. 

 As shown in Sec.~\ref{stiffEOS}, pairing tends to stiffen the equation of state in the high density regime.  Figure~\ref{fig:mu_con_H} shows the development of constituent quark masses and mean field pairing gaps; as $\mu$ increases, quark matter first appears in a 2SC phase in which only up and down quarks are paired ($\Delta_{ud} \neq 0$, $\Delta_{us} = \Delta_{ds} = 0$), and later evolves into a CFL phase in which all three quark flavors pair ($\Delta_{ud}, \Delta_{ds}, \Delta_{su} \neq 0$).  At $T = 0$ the 2SC-CFL transition appears to be first order for all NJL parameter sets.  However, given that this transition occurs at relatively low density ($n_B < 4 n_0$), the quark model results must be treated with caution.

Two-body correlations are also important at low densities. For example, in the constituent quark model, color magnetic interactions between quarks, in the presence of confinement, reduces the nucleon mass from three times the constituent quark mass, $\simeq 3\times 340 =1020$ MeV by some 80 MeV to its physical value.   Since confinement, by localizing the quarks into a spatial region $\sim \lqcd^{-3}$, increases the quark kinetic energies, as well as adding the energy of color flux tubes --
of typical length $\sim \lqcd^{-1}$ and energy $\sim \sigma \lqcd^{-1}$, where $\sigma$ is the string tension --
the energy gain from the color magnetic interaction must exceed $\sim 30$ MeV.

 The diquark interaction, $H$, treated in mean field, qualitatively simulates  the reduction in the average quark energy at low density  that results from pairing effects.    As the magnitude of the diquark interaction increases, the curves of the thermodynamic variables as functions of $\mu$ are shifted leftwards to lower chemical potential, as shown in Fig.~\ref{fig:mu_np_H}.  Thus, by including effects of pairing, one is able to maintain the stiff equation of state produced by a relatively large vector coupling, while at the same time enabling a smooth interpolation between the NJL and APR equations of state for all thermodynamic variables.

\begin{figure}
\includegraphics[width = 0.5\textwidth]{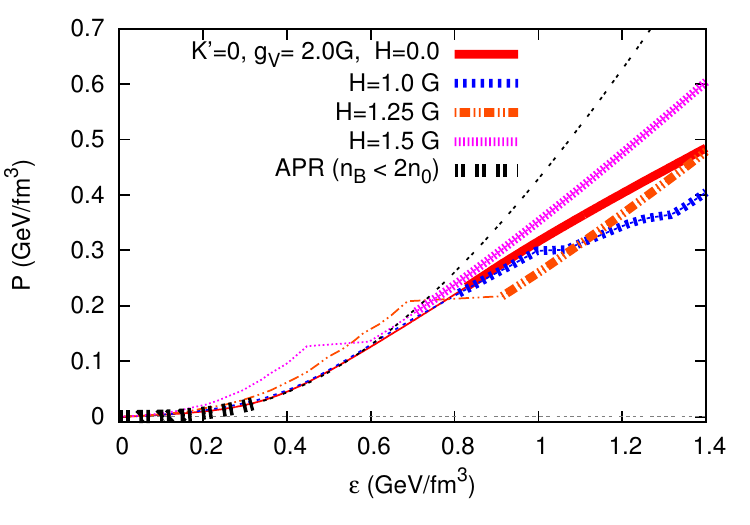}
\caption{\footnotesize{Pressure as a function of energy density for the same parameter sets as in Fig.~\ref{fig:mu_np_H}. 
The discontinuous change of $\varepsilon$ at fixed $P$ reflects the first order 2SC-CFL phase transition.
 For $0 < H < 1.5 G$, the equation of state is softened immediately following the 2SC-CFL phase transition, but as density increases, the equation of state eventually stiffens relative to the $H = 0$ case.  For $H=1.5G$, the equation of state is stiffer than the unpaired case for all densities. 
}}
\label{fig:ep_gV_H}
\vspace{-0.2cm}
\end{figure}

Figure~\ref{fig:ep_gV_H} demonstrates the impact of pairing on the stiffness of the NJL equation of state.  
The discontinuous change of $\varepsilon$ at fixed $P$ reflects the first order 2SC-CFL phase transition.
While for $0 < H < 1.5\, G$, the equations of state exhibit softening immediately following the 2SC-CFL transition, as the quark density increases further, pairing effects stiffen the equation of state for all NJL parameter sets, relative to the no-pairing case.  Moreover, when the pairing is sufficiently strong ($H > 1.5G$), the equation of state is stiffer than without pairing  ($H = 0$) across the entire density range.

  We note that for large $H$, the quark pressure at given $\mu$ exceeds the APR pressure even at very low densities.  Taken at face value, this would suggest that even at very low densities the ground state of QCD matter is quark rather than hadronic matter.  However, as we discuss in Sec.~\ref{sec:interpolatedEOS}, this high pressure is an unphysical consequence of the NJL model not being confining at low densities. 

\subsection{Chiral-diquark coupling: $K'$}

We now turn to the axial anomaly-induced coupling between the chiral and diquark condensates.  The importance of the anomalous coupling $K^\prime$ depends on the size of the chiral and diquark condensates. For $K^\prime >0$, this coupling favors the coexistence of chiral and diquark condensates \cite{Hatsuda:2006ps}.  Thus, as $K^\prime$ increases from zero the diquark condensate emerges at lower chemical potential and the chiral condensate persists to higher chemical potential.

\begin{figure}
\includegraphics[width = 0.5\textwidth]{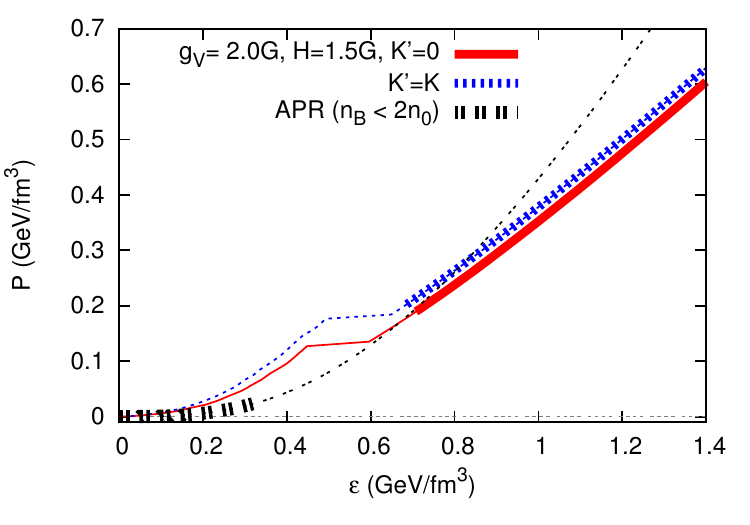}
\caption{\footnotesize{Pressure as a function of energy density in the presence of the chiral-diquark coupling, $K'=K$.  Larger $K'$ stiffens the equation of state, but its impact is significantly smaller than that of variations of $g_V$ or $H$.}}
\label{fig:ep_gV_H_K}
\vspace{-0.2cm}
\end{figure}

Figure~\ref{fig:ep_gV_H_K} shows the impact of $K^\prime$ on the NJL equation of state.  We note that while increasing $K^\prime$ from $0$ to $K$ slightly stiffens the equation of state, its impact is much smaller than that of $g_V$ or $H$.  Since the $K^\prime$ term in the Lagrangian 
can be read as a diquark interaction with an interaction strength proportional to the chiral condensate, the effect of $K^\prime$ can be largely absorbed by the variation of $g_V$ and $H$;  thus in the following we do not study the variation of $K'$ in detail but take $K^\prime = K$ as a canonical value.

\section{Interpolated EOS}
\label{sec:interpolatedEOS}

  We now discuss constructing an interpolated equation of state, $P (\mu)$ that smoothly joins the equations of state of the low density APR model to the high density NJL model.  The first step in defining an interpolation method for joining the hadronic and quark sectors is to determine the ``overlap region" in which the two equations of state will be merged.  Beginning in the low density hadronic regime described by APR, we expect that as density increases, corrections from many-body forces, hyperon degrees of freedom, multiple meson exchanges and the like, will become important above $n_B \sim 2 n_0$.   Thus, we fix the lower boundary of the interpolation to $n_< \equiv 2n_0$. 

 As the density decreases in the quark regime , confining effects, which trap quarks into baryons, become increasingly important.  Assuming that the radius of a typical baryon is $r_B \sim 0.4-0.5$ fm, baryons begin to percolate at around $n_B \sim 4-7 n_0$.  Thus, we set the upper boundary of the interpolation to $n_> \equiv 4-7 n_0$, with the precise location determined by the details of the given NJL parameter set being used. 

\begin{figure}
\includegraphics[width = 0.48\textwidth]{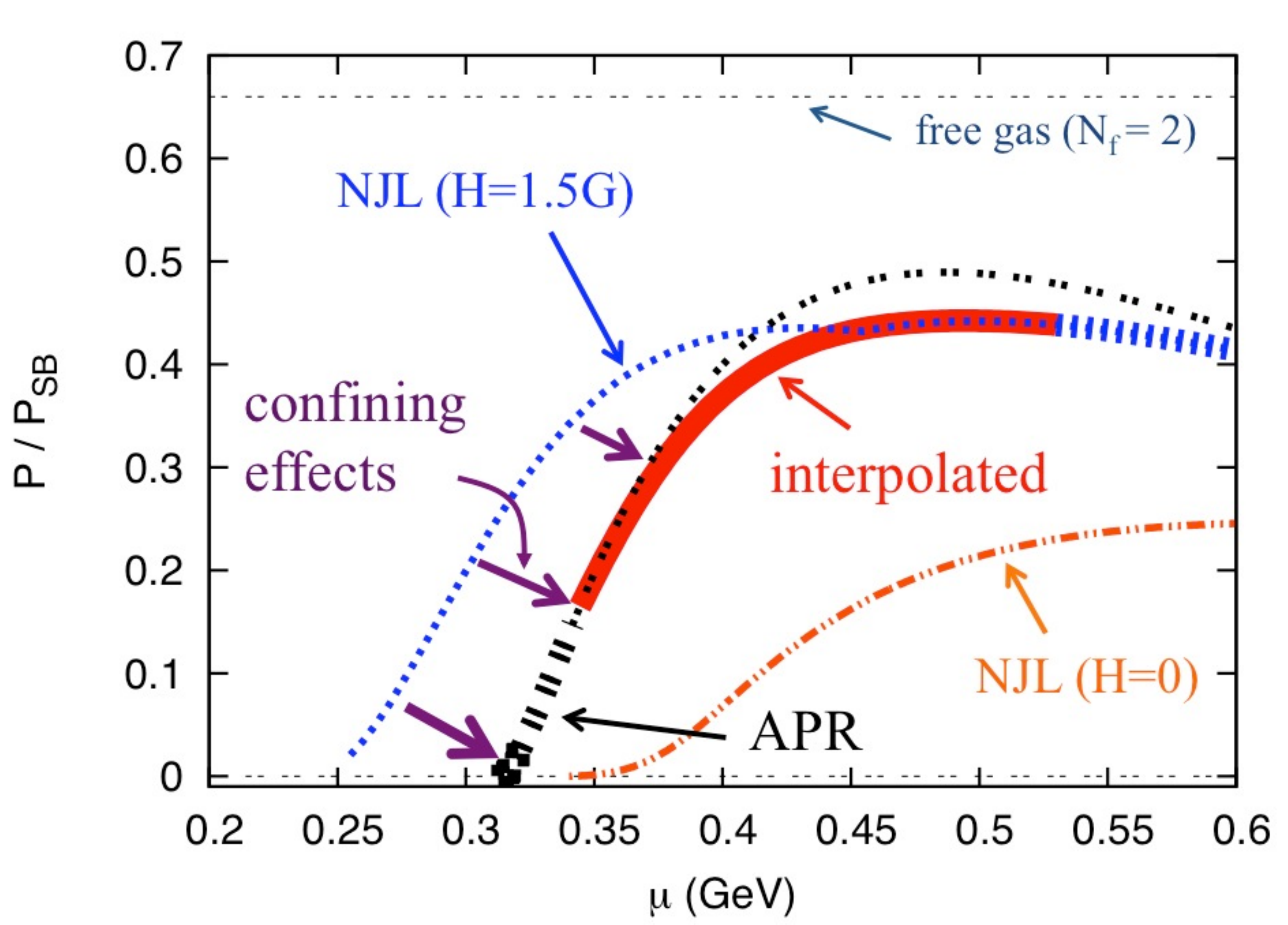}
\caption{\footnotesize{Schematic illustration of confining effects on the hybrid quark-hadron equation of state, $P (\mu)$, here normalized by the Stefan-Boltzman gas for $\Nf =3$. 
Effects of confinement are strong at lower density, suppressing the excess NJL pressure.  The boundaries of the interpolated equation of state are $(n_<, n_>) =(2.0,\, 5.0) n_0$.}}
\label{fig:confining_mu-p}
\vspace{-0.2cm}
\end{figure}

 In the intermediate density regime at a given chemical potential, the pressure of the interpolated equation of state should be lower than the NJL pressure extrapolated to lower densities (Fig.\ref{fig:confining_mu-p}).  This follows from the fact that nonconfining models yield excess quark populations at given chemical potential, due to the unphysically small energy cost of having quarks present.   In other words, were the confining effects of QCD incorporated in the description of the quark phase, the pressure, especially in dilute region, would be significantly suppressed.  This situation is quite analogous to the ``semi"-QGP picture for finite temperature QCD~\cite{Pisarski:2000eq} in which an ``overpopulated" quark pressure is suppressed by Polyakov loop effects, until thermal quarks and gluons exhibit quasiparticle behavior at temperatures beyond $\sim 2-3T_c$, where $T_c$ is the pseudocritical temperature for deconfinement ~\cite{Andersen:2011sf}.

   The present 3-window description is quite different from the conventional one involving a first order hadron-quark phase transition \cite{BC1976}.  In the latter case, the quark pressure at low density of nonconfining models must be smaller than the hadronic one, in order to ensure the intersection of the quark and hadronic pressure at reasonable density. This is achieved either by restricting the quark model parameters or by introducing a bag constant to lower the quark pressure. Such choices, however, generally affect the quark matter equation of state not only in the (presumably unreliable) low density limit, but also in the high density regime in which it should be reliable.
   
\subsection{Thermodynamic constraints}

Having discussed the qualitative aspects of a hadron-quark interpolation, we now briefly review the thermodynamic constraints imposed on this interpolation which are necessary to ensure that the interpolated equation of state is physical.  These constraints are as follows: (i) the pressure $P(\mu)$ must be continuous everywhere; (ii) $n_B(\mu) = (\partial P/\partial \mu)/\Nc$ must be a monotonically increasing function in order to ensure stability of the system with respect to phase separation: $\partial^2 P/\partial \mu^2 =  N_c\partial n_B/\partial \mu>0$.  (iii) In addition one physically expects that the speed of sound must be less than the speed of light: $c_s^2=\partial P/\partial \varepsilon <1$ \cite{ruderman-bludman,Ellis2007}.  These conditions tightly constrain possible interpolations of the equation of state. 

\subsection{Interpolation method}

    We now describe
a particular method for constructing a phenomenological quark-hadron equation of state.
To interpolate in the variables $\mu$-$P$, we employ a simple polynomial interpolation function, defining an $N$th order polynomial interpolant for the pressure:
\beq
\calP (\mu) = \sum^N_{m=0} b_m \mu^m \,,~~~ {\rm for}~\mu_< < \mu < \mu_>    ,
\label{interp}
\eeq
where $\mu_<$ and $\mu_>$ are defined as the points where $n_B(\mu_<) \equiv n_< =2n_0$ and $n_B(\mu_>) \equiv n_> =4-7 n_0$. The coefficients $b_m$ are chosen to satisfy the matching conditions at the boundaries of the interpolating interval.  At $\mu=\mu_<$:
\beq
\hspace{-0.5cm}
P_{{\rm APR}} (\mu_<) = \calP (\mu_<)\,,  
~\frac{\, \partial P_{ {\rm APR}} \,}{ \partial \mu }\bigg|_{\mu_<} = \frac{\, \partial \calP \,}{ \partial \mu }\bigg|_{\mu_<} \,,\cdots
\eeq
and at $\mu=\mu_>$:
\beq
\hspace{-0.5cm}
P_{{\rm NJL}} (\mu_>) = \calP (\mu_>)\,,  
~\frac{\, \partial P_{ {\rm NJL}} \,}{ \partial \mu }\bigg|_{\mu_>} = \frac{\, \partial \calP \,}{ \partial \mu }\bigg|_{\mu_>} \,,\cdots
\eeq
The number of derivatives that one matches at each boundary is a matter of choice.  In general, matching more derivatives results in a smoother interpolation, but at the same time increases the probability of producing unphysical artifacts in the interpolating region (e.g., inflection points in $P(\mu)$ which violate thermodynamic constraints).  Here we match up to second order derivatives at each boundary, which ensures that the pressure, number density, and number susceptibility are continuous.  Correspondingly, these six boundary conditions require that Eq.~(\ref{interp}) has six terms ($N = 5)$.

\begin{figure}
\includegraphics[width = 0.48\textwidth]{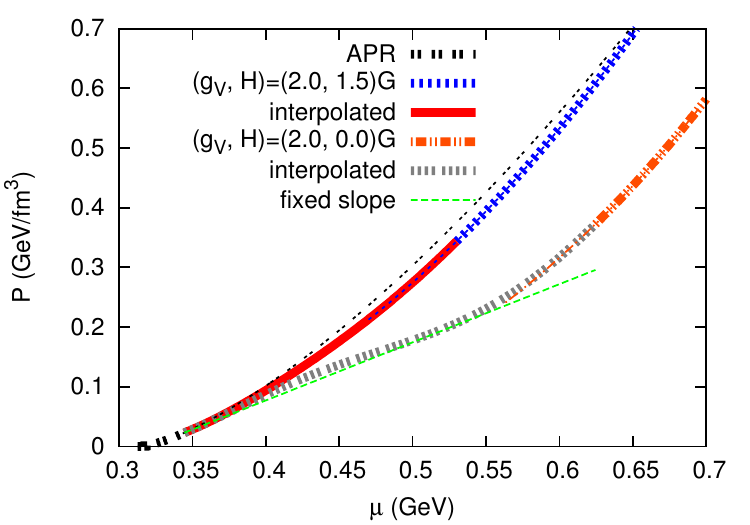}
\caption{\footnotesize{Interpolated equations of state, $P (\mu)$, for two parameter sets, with up to second order derivatives matched at the interpolation boundaries.  As a guide to the eye, we also plot the pressure with constant slope fixed at $n_<$.  Thermodynamic stability requires that the slopes in the interpolated equation of state increase monotonically: $\partial^2P/\partial \mu^2 =  \Nc \partial n_B/\partial \mu >0$, which is violated for the equation of state constructed in the simple interpolation with parameter Set II.}}
\label{fig:inter6mu-p}
\vspace{-0.2cm}
\end{figure}

\begin{figure}
\includegraphics[width = 0.47\textwidth]{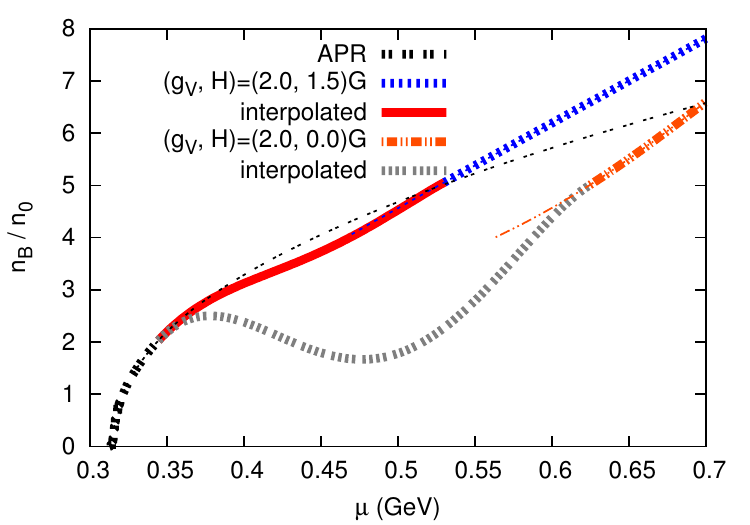}
\caption{\footnotesize{Baryon density as a function of chemical potential for the interpolated equations of state.  The parameter set $(g_V, H)=(2.0, 0.0)G$ has an unstable region $\partial n_B/\partial \mu <0$ with the simple six-term interpolating function.}}
\label{fig:inter6mu-n}
\vspace{-0.2cm}
\end{figure}

\subsection{Interpolated EOS}

Figure~\ref{fig:inter6mu-p} shows the interpolated equation of state $P (\mu)$, with the interpolation boundaries $(n_<, n_>)=(2.0, 5.0) n_0$.  For illustration, we consider two NJL parameter sets: 
\beq
(g_V, H)=\left\{
\begin{matrix}
&(2.0,\, 1.5)\, G ~~~~({\rm Set\, I})\\
&\,\,(2.0,\, 0.0)\, G~~~~({\rm Set\, II})
\end{matrix}\right.
\eeq 
with $K' = K$ in both cases.  For $n_> \simeq 4 n_B$, Set I satisfies all the conditions demanded by the thermodynamic constraints, as we verify shortly.  One cannot, however, within the present polynomial interpolation, construct a sensible interpolation for Set II, because at the interpolation boundaries the APR and NJL pressures are rather widely separated in $\mu$, a possibility noted in Sec.~\ref{sub:H}.  This wide separation in $\mu$ requires a small slope of the interpolated pressure, but at the same time the slope must be larger than the slope of the APR pressure at the lower boundary, because of the compressibility condition $\partial^2P/\partial \mu^2 >0$.  The interpolated equation of state has a region of $\partial n_B/\partial \mu <0$, as is clearly seen in the plot of $n_B (\mu)$ in Fig.~\ref{fig:inter6mu-n}. 

 The result presented in Fig.~\ref{fig:inter6mu-n} does not preclude constructing a sensible interpolated pressure for Set II.   As one sees in  Fig.~\ref{fig:inter6mu-n}, it is possible to join the low and high baryon density curves with a nondecreasing function of $\mu$. The present exercise shows that the class of interpolating functions for Set II is much more restrictive than for Set I.  For example, if there is a first order phase transition between the hadronic and quark regions, the possible density discontinuities are smaller for Set II than Set I.  More detailed treatments of the interpolation region are beyond the scope of the present work and we henceforth restrict our consideration to the simple polynomial interpolation, rejecting NJL parameter sets, such as II, incapable of being joined with APR in a thermodynamically consistent manner.
 
\begin{figure}
\includegraphics[width = 0.45\textwidth]{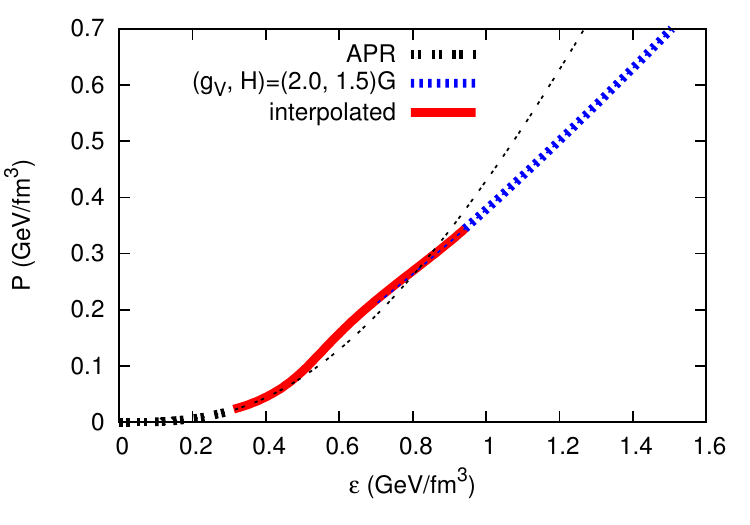}
\vspace{-0.2cm}
\caption{\footnotesize{Pressure vs. energy density for the interpolated equation of state for $(g_V, H)=(2.0, 0.0)G$, $K'=K$, and $(n_<, n_>)=(2.0, 5.0) n_0$.}}
\label{fig:inter6e-p}
\vspace{-0.2cm}
\end{figure}

\begin{figure}
\includegraphics[width = 0.45\textwidth]{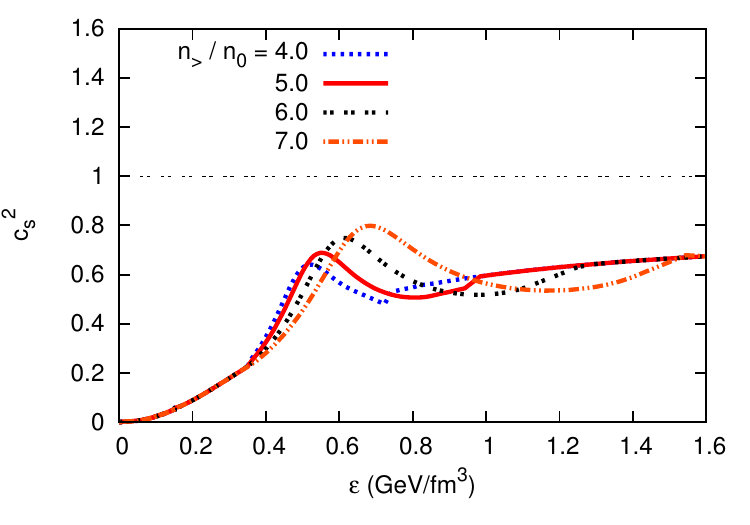}
\vspace{-0.2cm}
\caption{\footnotesize{The speed of sound $c_s^2=\partial P/\partial \varepsilon$ as a function of $\varepsilon$. The NJL parameter set is the same as Fig.~\ref{fig:inter6e-p}, for different values of the high density boundary: $n_>/n_0$ = 4.0, 5.0, 6.0, and 7.0.  For each of these choices, the causality condition is satisfied.}}
\label{fig:inter6cs}
\vspace{-0.2cm}
\end{figure}

Figure~\ref{fig:inter6e-p} shows the pressure vs. energy density for parameter set I.  In this case, the high density equation of state is as stiff as APR extrapolated into the region beyond $n_B=2n_0$.  From Fig.~\ref{fig:inter6cs}, we observe that the causality constraint is satisfied when the high density boundary, $n_>$, is varied from $4 n_0$ to $7 n_0$. If we take $n_> \lesssim 4 n_0$, however, we find that $c_s^2 > 1$, a putative violation of causality.  

Finally, we consider the possible impact of a gluonic bag constant, $B_g \sim (200 {\rm MeV})^4\simeq 0.2\, {\rm GeV fm}^{-3}$, which should be included when the gluon sector becomes perturbative.   This contribution reduces the pressure in the quark matter region by $30-40$\%, which makes it extremely difficult to interpolate between the hadronic and quark regimes  without violating the condition $\partial n_B/\partial \mu>0$ (cf. Fig. \ref{fig:inter6mu-p}).  At the same time, the bag constant increases the energy density by $B_g$, so the resulting equation of state becomes significantly softer.  Strictly speaking, even in this situation it would be possible to construct equations of state by increasing $g_V$ and $H$ significantly from our current choices;  however the current choices for these couplings are already relatively large and it is difficult to identify a mechanism that would significantly increase either coupling in the dense regime.  Thus, we conclude that $B_g$ should be very small in the quark matter equation of state, even at $n_B\sim 10n_0$; the gluons remain condensed and nonperturbative.

\section{Mass-radius relation}
\label{sec:MR}

\begin{figure}
\includegraphics[width = 0.48\textwidth]{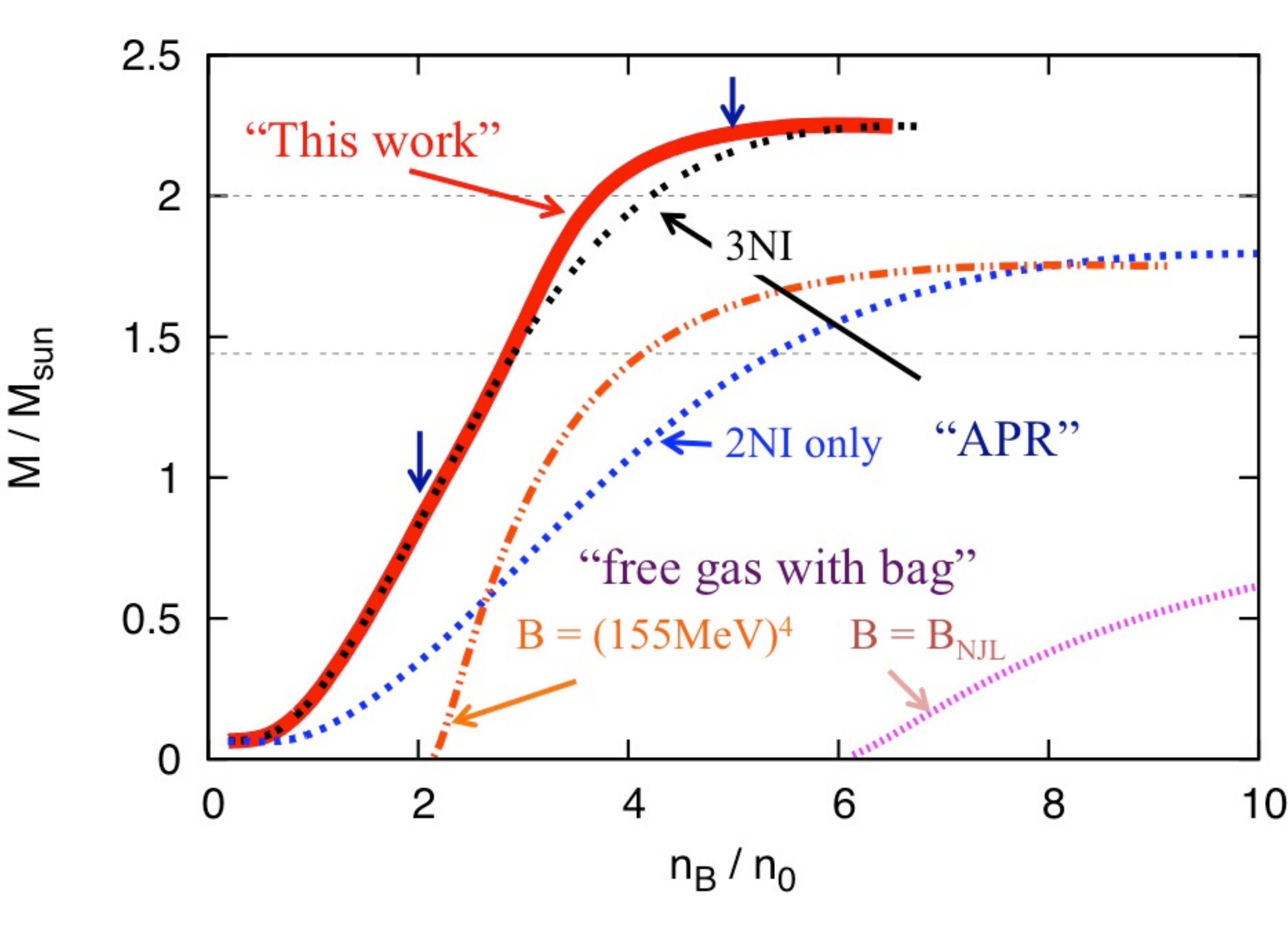}
\caption{\footnotesize{Neutron star mass as a function of central baryon density, $n_B^c$.  We display curves corresponding to our interpolated equation of state for $(g_V,H,K')=(2G, 1.5G, K)$ and $n_>=5 n_0$, as well as APR with and without the three-nucleon interaction, and the three-flavor free quark gas with bag constants $B_q=B_{ {\rm NJL} }=(219 {\rm MeV})^4$ and $B_q=(155{\rm MeV})^4$. The vertical arrows indicate the boundaries of the interpolation region for the interpolated equation of state. }}
\label{fig:M-n}
\vspace{-0.2cm}
\end{figure}

By solving the TOV equation for a given value of the baryon density at the center, we construct a family of stars whose masses and radii are functions of $n^c_B$.
		The $M$-$n_B^c$ relation for the equation of state with NJL parameters $(g_V, H, K^\prime) = (2G, 1.5G, K)$ is shown in Fig.~\ref{fig:M-n}, where we have taken $n_>=5n_0$ as in the previous section.  To examine effects of the quark bag constant associated with chiral restoration, we also show results for the three-flavor free quark gas with bag constants $B_q=B_{ {\rm NJL} }=(219 {\rm MeV})^4$ and $B_q=(155{\rm MeV})^4$. In the NJL model by itself, the quark bag constant is so large that neutron star masses are restricted to $M < M_\odot$.  However, we note that as the bag constant decreases, neutron star masses rise, so that models yielding smaller bag constants allow more massive stars.  

We see in Fig.~\ref{fig:M-n} that the $M (n^c_B)$ curves for our interpolated equation of state and APR are quite similar, a not too surprising result given that our chosen NJL parameter set yields an equation of state quite similar to APR.  However, while the thermodynamic properties of the two systems are similar, the underlying effective degrees of freedom are quite different.  Indeed, APR is well known for its extreme stiffness at high densities, but the effect of hyperonic degrees of freedom in the hadronic sector are expected to reduce this stiffness.
On the other hand, our NJL treatment of the quark sector includes strange quarks from the beginning, and is capable of producing a sufficiently stiff equation of state to support stars whose masses exceed $2M_\odot$ \cite{Masuda:2012kf}. 

\begin{figure}
\includegraphics[width = 0.48\textwidth]{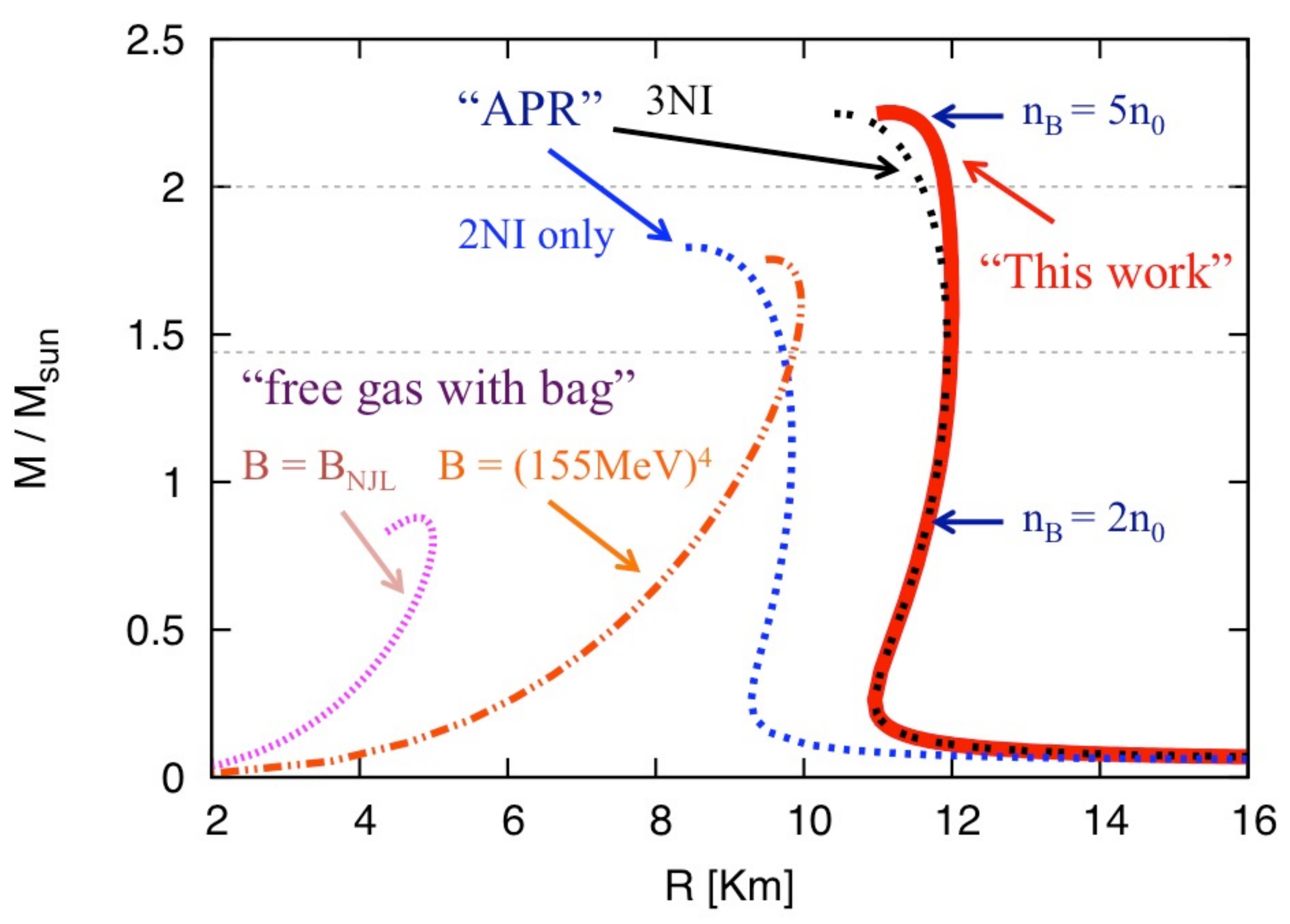}
\caption{\footnotesize{Mass-radius relation for neutron stars with several equations of state (same as Fig.~\ref{fig:M-n}). }}
\label{fig:M-R}
\vspace{-0.2cm}
\end{figure}

  Figure~\ref{fig:M-R} shows the $M$-$R$ relation for the interpolated equation of state with parameter set I.   For most central densities the stellar radius is $\sim 11-12$ km, which is compatible with observational data~ \cite{{Ozel2010,Steiner2012}}. However, the radius at which the mass starts to rise in the $M$-$R$ plot is sensitive to the properties of the hadronic equation of state for $n_0 \sim 1.0-2.0 n_0$, with different hadronic equations of state yielding radii differing by up to $\sim 2$ km \cite{Masuda:2012kf}.   

\section{Summary}
\label{sec:summary}

In this paper we have constructed a phenomenological equation of state over the range of baryon densities $n_B \sim 1-10 n_0$ by interpolating between the low density hadronic APR equation of state and the high density NJL quark model.  In so doing, we explored a number of relevant ingredients of the equation of state needed to realize neutron stars of mass $\sim 2M_\odot$, while satisfying necessary thermodynamic and causality constraints.  These requirements constrain the form of the interpolated equation of state and allow one to infer qualitative effects regarding the intermediate density region between the hadronic and quark regimes.  The repulsive density-density interaction, color-magnetic interaction, and confining effects play a vital role in determining the structure of the equation of state.  A crucial result is that the gluonic bag constant must be small, i.e., the gluons must remain strongly coupled throughout the density region of interest in massive neutron stars; the gluon sector remains nonperturbative even at $\sim 10n_0$.    One reaches a similar conclusion from studies of quarkyonic matter \cite{McLerran:2007qj,Kojo:2011fh,Fukushima:2014pha}, in which nonperturbative gluons in quark matter play a crucial role. 

We have also emphasized why the three window model of interpolation~\cite{Masuda:2012kf} is capable of producing stiffer equations of state than conventional hybrid equations of state involving first order phase transitions.  As opposed to conventional models, which require the intersection of quark and hadronic $P(\mu)$ curves, we propose that the quark pressure $P(\mu)$ based on nonconfining models need not (and even \textit{should} not) intersect the hadronic equation of state before the inclusion of confining effects. While this observation is not directly applicable at low density (where we did not use the NJL model), it results in a much wider range of possible high density quark equations of state. As a result, we are able to explore a region of parameter space that has been omitted from prior studies, while still producing a stiff equation of state required to support massive neutron stars.

  In this work we have not taken into account possible meson condensed phases, by which we mean condensates in which the order parameter has the quantum number of a mesonic field.   Such condensates have been studied in a nuclear context~\cite{Migdal:1978az,Baym-picond,Kaplan:1986yq} as well in quark matter, e.g., 
 inhomogeneous diquark~\cite{Alford:2007xm} and chiral~\cite{Buballa:2014tba} condensates.  If extant, such exotic phases would likely occur in the the neither purely hadronic nor purely quark density region in which we have interpolated.  Thus one cannot directly take over previous results for meson condensates, including the strength, or density discontinuity, of the first order phase transition to the condensed phase.   For a given hadronic equation of state at low density and quark equation of state at high density, the strength of such a phase transition is bounded.
Although we have, for simplicity, considered only a smooth interpolation scheme, one should more generally allow for such exotic phases; then the smooth $P(\mu)$ curve used in this work would be replaced by one with a small kink, keeping the positive curvature of $P(\mu)$ in the interpolation.
We anticipate that even with condensates at intermediate density, it will still be possible to find a reasonable parameter set for the color-magnetic and vector interactions that is compatible with the existence of massive stars. The issue of exotic phases remains open until we can reliably estimate the high density quark equations of state. Further studies are needed to understand better the impact of exotic phases on the in-medium NJL parameters and in turn, their implications for the description of massive neutron stars.

In order to improve the description of the intermediate density matter in neutron stars it is important to further refine our understanding of the hadronic equation of state near $n_B \sim (1-3) n_0$.  In particular, a more careful assessment of the importance of many-body interactions and the emergence of hyperons is required.  Further constraints may be obtained from heavy-ion collisions, including strangeness production~\cite{Klahn:2006ir}, and lattice QCD calculations of hyperon-nucleon interactions~\cite{Aoki:2012tk}. Studies of the density dependence of nuclear forces in terms of quarks and gluons play a crucial role in determining when (and why) quarks emerge as the proper degrees of freedom at high density. It would be desirable in the future to extend to finite temperature the present approach to the equation of state to enable us to address dynamical questions such as applications to heavy ion collisions, and neutron star cooling. 
It is also important to obtain an improved estimate of the quark bag constant, for were it much smaller than the NJL estimate used here, the softening associated with chiral restoration would be significantly reduced and large vector and diquark couplings would not be necessary to obtain a stiff quark matter equation of state within the present context.

\section{Acknowledgements}
Author T.K. thanks Y. Hidaka, A. Ohnishi, and R. Pisarski for enlightening discussions during the Fourth APS-JPS joint meeting HAWAII 2014, and M. Alford, S. Han, and K. Schwenzer for discussions during his visit to Washington University.  He is also grateful to D. Blaschke for discussions and his lectures given at the University of Bielefeld in 2013. 
  Authors G.B. and T.K. thank T. Hatsuda and K. Masuda for numerous discussions of the  equation of state at intermediate densities. This research was supported in part by NSF Grants PHY09-69790 and PHY13-05891.



\begin{thebibliography}{99}

\bibitem{BC1976}  G. Baym and S. A. Chin,  Phys. Lett. 62B, 241 (1976).

\bibitem{Lattimer:2006xb}
Reviewed in
  J.~M.~Lattimer and M.~Prakash,
  Phys.\ Rept.\  {\bf 442} (2007) 109
  [astro-ph/0612440].

\bibitem{Fukushima:2010bq}
  K.~Fukushima and T.~Hatsuda,
  Rept.\ Prog.\ Phys.\  {\bf 74} (2011) 014001
  [arXiv:1005.4814 [hep-ph]];
  K.~Fukushima and C.~Sasaki,
  Prog.\ Part.\ Nucl.\ Phys.\  {\bf 72}, 99 (2013) 
  [arXiv:1301.6377 [hep-ph]].
\bibitem{Alford:2007xm}
For a review of the color superconductivity,
  M.~G.~Alford, A.~Schmitt, K.~Rajagopal and T.~Schäfer,
  Rev.\ Mod.\ Phys.\  {\bf 80} (2008) 1455
  [arXiv:0709.4635 [hep-ph]].
\bibitem{Buballa:2014tba}
For a review of the inhomogeneous chiral phases,
  M.~Buballa and S.~Carignano,
  arXiv:1406.1367 [hep-ph].




\bibitem{Demorest}P. Demorest, T. Pennucci, S. M. Ransom, R. M. S. Roberts, and J. W. T. Hessels, Nature (London) \textbf{467}, 1081 (2010).
\bibitem{Antoniadis2013}J. Antoniadis, {\it et al}., Science \textbf{340}, 1233232 (2013)
\bibitem{Romani2012}R. W. Romani, A. V. Filippenko, J. M. Silverman, S. B. Cenko, J. Greiner, A. Rau, J. Elliott, and H. J. Pletsch, Astrophys. J. Lett. \textbf{760}, L36 (2012). 

\bibitem{Ozel2010}F. \"{O}zel, G. Baym, and T. G\"{u}ver, Phys. Rev. D \textbf{82} 010301 (2010).

\bibitem{Steiner2012}A. W. Steiner, J. M. Lattimer, and E. F. Brown,  Astrophys. J, {\bf 722}  33, (2010); 
Astrophys. J. Lett. \textbf{765}, L5 (2013).

  
\bibitem{ruderman-bludman} M. A. Ruderman and S. A. Bludman,  Phys. Rev. {\bf 170}, 1176 (1968).

\bibitem{Ellis2007} G. F. R. Ellis, R. Maartens, and M. A. H. MacCallum, Gen. Relativ. Gravit. \textbf{39}, 1651 (2007).

\bibitem{Weinberg:1978kz}
  S.~Weinberg,
  Physica A {\bf 96} (1979) 327;
  {\it ibid.} Nucl.\ Phys.\ B {\bf 363} (1991) 3;
  {\it ibid.} Phys.\ Lett.\ B {\bf 251} (1990) 288.
  
  
\bibitem{Epelbaum:2008ga}
  E.~Epelbaum, H.~W.~Hammer and U.~G.~Meissner,
  Rev.\ Mod.\ Phys.\  {\bf 81} (2009) 1773
  [arXiv:0811.1338 [nucl-th]];
  E.~Epelbaum,
  Prog.\ Part.\ Nucl.\ Phys.\  {\bf 57} (2006) 654
  [nucl-th/0509032].
\bibitem{Bogner:2009bt}
  S.~K.~Bogner, R.~J.~Furnstahl and A.~Schwenk,
  Prog.\ Part.\ Nucl.\ Phys.\  {\bf 65} (2010) 94
  [arXiv:0912.3688 [nucl-th]].
 
\bibitem{Akmal:1998cf}
  A.~Akmal, V.~R.~Pandharipande and D.~G.~Ravenhall,
  Phys.\ Rev.\ C {\bf 58} (1998) 1804
  [nucl-th/9804027].


\bibitem{Freedman:1976ub}
  B.~A.~Freedman and L.~D.~McLerran,
  Phys.\ Rev.\ D {\bf 16} (1977) 1169;
  {\it ibid.} {\bf 17} (1978) 1109.
\bibitem{Kurkela:2009gj}
  A.~Kurkela, P.~Romatschke and A.~Vuorinen,
  Phys.\ Rev.\ D {\bf 81} (2010) 105021
  [arXiv:0912.1856 [hep-ph]].
\bibitem{Fraga:2001id}
  E.~S.~Fraga, R.~D.~Pisarski and J.~Schaffner-Bielich,
  Phys.\ Rev.\ D {\bf 63} (2001) 121702
  [hep-ph/0101143];
  E.~S.~Fraga, A.~Kurkela and A.~Vuorinen,
  Astrophys.\ J.\ Lett. {\bf 781}, 2, L25 (2014)
  [arXiv:1311.5154 [nucl-th]];
  A.~Kurkela, E.~S.~Fraga, J.~Schaffner-Bielich and A.~Vuorinen,
  Astrophys.\ J.\  {\bf 789} (2014) 127
  [arXiv:1402.6618 [astro-ph.HE]].
  

\bibitem{Masuda:2012kf}
  K.~Masuda, T.~Hatsuda and T.~Takatsuka,
  Astrophys.\ J.\  {\bf 764}, 12 (2013)
  [arXiv:1205.3621 [nucl-th]];
Prog. Theor. Exp. Phys. (2013) 073D01.
  [arXiv:1212.6803 [nucl-th]].
  
\bibitem{perc} G. Baym,  Physica (Amsterdam) {\bf 96A}, 131 (1979).

\bibitem{Kunihiro:1991qu}
  T.~Kunihiro,
  Phys.\ Lett.\ B {\bf 271} (1991) 395.

\bibitem{De Rujula:1975ge}
  A.~De Rujula, H.~Georgi and S.~L.~Glashow,
  Phys.\ Rev.\ D {\bf 12} (1975) 147.

\bibitem{Asakawa:1989bq}
  M.~Asakawa and K.~Yazaki,
  Nucl.\ Phys.\ A {\bf 504} (1989) 668.
  
 



\bibitem{Hatsuda:1994pi}
  T.~Hatsuda and T.~Kunihiro,
  Phys.\ Rept.\  {\bf 247} (1994) 221
  [hep-ph/9401310];
\bibitem{Buballa:2003qv}
  M.~Buballa,
  Phys.\ Rept.\  {\bf 407} (2005) 205
  [hep-ph/0402234].



\bibitem{McLerran:2007qj}
  L.~McLerran and R.~D.~Pisarski,
  Nucl.\ Phys.\ A {\bf 796} (2007) 83
  [arXiv:0706.2191 [hep-ph]].


\bibitem{Alford:2002rj}
  M.~Alford and S.~Reddy,
  Phys.\ Rev.\ D {\bf 67} (2003) 074024
  [nucl-th/0211046].
\bibitem{Alford:2004pf}
  M.~Alford, M.~Braby, M.~W.~Paris and S.~Reddy,
  Astrophys.\ J.\  {\bf 629} (2005) 969
  [nucl-th/0411016].

\bibitem{Klahn:2006iw}
  T.~Klahn, D.~Blaschke, F.~Sandin, C.~Fuchs, A.~Faessler, H.~Grigorian, G.~Ropke and J.~Trumper,
  Phys.\ Lett.\ B {\bf 654} (2007) 170
  [nucl-th/0609067].
\bibitem{Bonanno:2011ch}
  L.~Bonanno and A.~Sedrakian,
  Astron.\ Astrophys.\  {\bf 539} (2012) A16
  [arXiv:1108.0559 [astro-ph.SR]].




\bibitem{Kitazawa:2002bc}
  M.~Kitazawa, T.~Koide, T.~Kunihiro and Y.~Nemoto,
  Prog.\ Theor.\ Phys.\  {\bf 108} (2002) 929
  [hep-ph/0207255, hep-ph/0307278].


\bibitem{Kobayashi:1970ji}
  M.~Kobayashi and T.~Maskawa,
  Prog.\ Theor.\ Phys.\  {\bf 44} (1970) 1422;
  G.~'t Hooft,
  Phys.\ Rev.\ Lett.\  {\bf 37} (1976) 8;
{\it ibid.} 
  Phys.\ Rev.\ D {\bf 14} (1976) 3432
   [Erratum-ibid.\ D {\bf 18} (1978) 2199].

\bibitem{Hatsuda:2006ps}
  T.~Hatsuda, M.~Tachibana, N.~Yamamoto and G.~Baym,
  Phys.\ Rev.\ Lett.\  {\bf 97} (2006) 122001 [hep-ph/0605018];
  H.~Abuki, G.~Baym, T.~Hatsuda and N.~Yamamoto,
  Phys.\ Rev.\ D {\bf 81} (2010) 125010 [arXiv:1003.0408 [hep-ph]];
  P.~D.~Powell and G.~Baym,
  Phys.\ Rev.\ D {\bf 88} (2013) 1,  014012 [arXiv:1302.0416 [hep-ph]].


\bibitem{Iida:2000ha}
  K.~Iida and G.~Baym,
  Phys.\ Rev.\ D {\bf 63} (2001) 074018
   [Erratum-ibid.\ D {\bf 66} (2002) 059903]
  [hep-ph/0011229];
  M.~Alford and K.~Rajagopal,
  JHEP {\bf 0206} (2002) 031
  [hep-ph/0204001];
  A.~W.~Steiner, S.~Reddy and M.~Prakash,
  Phys.\ Rev.\ D {\bf 66} (2002) 094007
  [hep-ph/0205201].
 
\bibitem{Buballa:2005bv}
  M.~Buballa and I.~A.~Shovkovy,
  Phys.\ Rev.\ D {\bf 72} (2005) 097501
  [hep-ph/0508197].
  
\bibitem{Abuki:2005ms}
  H.~Abuki and T.~Kunihiro,
  Nucl.\ Phys.\ A {\bf 768} (2006) 118
  [hep-ph/0509172].


\bibitem{Blaschke:2005uj}
  D.~Blaschke, S.~Fredriksson, H.~Grigorian, A.~M.~Oztas and F.~Sandin,
  Phys.\ Rev.\ D {\bf 72} (2005) 065020
  [hep-ph/0503194].
  

\bibitem{Rehberg}P. Rehberg, S. P. Klevansky, and J. Hefner, Phys. Rev. C \textbf{53}, 410 (1996).

\bibitem{Lutz1992}M. Lutz, S. Klimt, and W. Weise, Nucl. Phys. A \textbf{542}, 521 (1992).

\bibitem{Witten:1984rs}
  E.~Witten,
  Phys.\ Rev.\ D {\bf 30} (1984) 272.

\bibitem{Book}
Kohsuke Yagi, Tetsuo Hatsuda, and Yasuo Miake,
{\it Quark-Gluon Plasma: From Big Bang to Little Bang}, Cambridge Monographs on Particle Physics, Nuclear Physics and Cosmology
 (Cambridge University, Cambridge, England, 2008)
 
\bibitem{Walecka:1974qa}
  J.~D.~Walecka,
  Annals Phys.\  {\bf 83} (1974) 491;
  B.~D.~Serot and J.~D.~Walecka,
  Adv.\ Nucl.\ Phys.\  {\bf 16} (1986) 1;
{\it ibid.}, Int.\ J.\ Mod.\ Phys.\ E {\bf 06} (1997) 515
  [nucl-th/9701058].

 
\bibitem{Bratovic:2012qs}
  N.~M.~Bratovic, T.~Hatsuda and W.~Weise,
  Phys.\ Lett.\ B {\bf 719} (2013) 131
  [arXiv:1204.3788 [hep-ph]].
\bibitem{Kaczmarek:2011zz}
  O.~Kaczmarek, F.~Karsch, E.~Laermann, C.~Miao, S.~Mukherjee, P.~Petreczky, C.~Schmidt and W.~Soeldner {\it et al.},
  Phys.\ Rev.\ D {\bf 83} (2011) 014504
  [arXiv:1011.3130 [hep-lat]].


\bibitem{Pisarski:2000eq}
  R.~D.~Pisarski,
  Phys.\ Rev.\ D {\bf 62} (2000) 111501
  [hep-ph/0006205];
  A.~Dumitru, Y.~Guo, Y.~Hidaka, C.~P.~K.~Altes and R.~D.~Pisarski,
  Phys.\ Rev.\ D {\bf 83} (2011) 034022
  [arXiv:1011.3820 [hep-ph]];
  K.~Fukushima,
  Phys.\ Lett.\ B {\bf 591} (2004) 277
  [hep-ph/0310121].

\bibitem{Andersen:2011sf}
  J.~O.~Andersen, L.~E.~Leganger, M.~Strickland and N.~Su,
  JHEP {\bf 1108} (2011) 053
  [arXiv:1103.2528 [hep-ph]];
  J.~O.~Andersen, S.~Mogliacci, N.~Su and A.~Vuorinen,
  Phys.\ Rev.\ D {\bf 87} (2013) 074003
  [arXiv:1210.0912 [hep-ph]];
  N.~Haque, A.~Bandyopadhyay, J.~O.~Andersen, M.~G.~Mustafa, M.~Strickland and N.~Su,
  JHEP {\bf 1405} (2014) 027
  [arXiv:1402.6907 [hep-ph]].

\bibitem{Kojo:2011fh}
  T.~Kojo,
  Nucl.\ Phys.\ A {\bf 877} (2012) 70
  [arXiv:1106.2187 [hep-ph]].
\bibitem{Fukushima:2014pha}
  K.~Fukushima, Nucl. Phys. {\bf A931}, 257 (2014).
  arXiv:1408.0547 [hep-ph].

\bibitem{Migdal:1978az}
  A.~B.~Migdal,
  Rev.\ Mod.\ Phys.\  {\bf 50} (1978) 107.

\bibitem{Baym-picond} G. Baym,  Phys. Rev.  Lett. {\bf 30}, 1340 (1973); G. Baym and D. Campbell,  in
{\it Mesons in Nuclei} edited by M. Rho and D. Wilkinson (North Holland, Amsterdam, 1978),
Vol.  III, pp. 1031.


\bibitem{Kaplan:1986yq}
  D.~B.~Kaplan and A.~E.~Nelson,
  Phys.\ Lett.\ B {\bf 175} (1986) 57.

\bibitem{Klahn:2006ir}
  T.~Klahn, D.~Blaschke, S.~Typel, E.~N.~E.~van Dalen, A.~Faessler, C.~Fuchs, T.~Gaitanos and H.~Grigorian {\it et al.},
  Phys.\ Rev.\ C {\bf 74} (2006) 035802
  [nucl-th/0602038].

\bibitem{Aoki:2012tk}
  S.~Aoki {\it et al.}  [HAL QCD Collaboration],
  PTEP (2012) 01A105
  [arXiv:1206.5088 [hep-lat]];
  T.~Doi {\it et al.}  [HAL QCD Collaboration],
  Prog.\ Theor.\ Phys.\  {\bf 127} (2012) 723
  [arXiv:1106.2276 [hep-lat]].

\end{thebibliography}
\end{document}